%% file: elsarticle-template.tex
\documentclass[preprint,10pt]{elsarticle}

\usepackage{amssymb}

\usepackage{lineno}

\usepackage{color}

\definecolor{Ora}{cmyk}{1, 0, 0, 0}

\input{preamble}

\journal{Applied Energy}

\begin{document}

\begin{frontmatter}

\title{{A Statistical Framework for District Energy Long-term Electric Load 
Forecasting}}

\author[address1]{Emily Royal}
\address[address1]{%
Colorado School of Mines, Department of Electrical Engineering, 1500 Illinois Street, Golden, CO 80401}
\ead{royal@mines.edu}
\cortext[cor1]{Corresponding author}

\author[address2]{Soutir Bandyopadhyay} 
\address[address2]{%
Colorado School of Mines, Department of Applied Mathematics and Statistics, 1500 Illinois Street, Golden, CO 80401} 
\ead{sbandyopadhyay@mines.edu}

\author[address3]{Alexandra Newman\corref{cor1}} 
\address[address3]{%
Colorado School of Mines, Department of Mechanical Engineering, 1500 Illinois Street, Golden, CO 80401} 
\ead{anewman@mines.edu}

\author[address1]{Qiuhua Huang} 
\ead{qiuhuahuang@mines.edu}

\author[address3]{Paulo Cesar Tabares-Velasco} 
\ead{tabares@mines.edu}
\begin{abstract}
\input{abstract}
\end{abstract}


\end{frontmatter}


\renewcommand{\arraystretch}{1.0}

\input{intro}

\input{lit_review}

\input{methodology}

\input{results_test_case}

\input{conclusion}

\input{acknowledgements}

\appendix
\setcounter{figure}{0} \renewcommand{\thefigure}{A.\arabic{figure}}
\setcounter{table}{0} \renewcommand{\thetable}{A.\arabic{table}}
\input{appendix}

\setcounter{figure}{0} \renewcommand{\thefigure}{B.\arabic{figure}}
\setcounter{table}{0} \renewcommand{\thetable}{B.\arabic{table}}
\input{appendixB}

\clearpage


\bibliographystyle{elsarticle-num-names}

\bibliography{forecast}

\biboptions{sort&compress}
\end{document}

%% file: preamble.tex
\usepackage{amsmath,amssymb,wasysym}  

 \usepackage{booktabs}

\usepackage[usenames,dvipsnames]{xcolor}	
\usepackage{bbm}	
\usepackage{subfig}

\usepackage{enumitem} 	

\usepackage{placeins}
\usepackage{longtable}	
\usepackage{multirow}	

\usepackage{colortbl}		

\usepackage{xifthen}		
\usepackage{xspace}		

\usepackage{nomencl}
\makenomenclature
\usepackage{xspace}
\usepackage{tabularx}  

\usepackage{amsmath,epsf,amssymb}
\usepackage{setspace}
\usepackage{color}
\usepackage{lineno}
\usepackage[usenames,dvipsnames]{xcolor}
\usepackage{fancyvrb}
\usepackage{url}
\usepackage{wasysym}
\usepackage{longtable}
\usepackage{lipsum}
\usepackage{adjustbox}
\usepackage[acronym]{glossaries}
\usepackage [english]{babel}
\usepackage{enumitem}
\usepackage{indentfirst}
\usepackage{relsize}
\usepackage{relsize}
\usepackage{cancel}
\usepackage[normalem]{ulem}

\newif\ifsrok
\srokfalse
\sroktrue

\graphicspath{{./figures/}}    
\usepackage{multirow}	
\usepackage{amssymb,wasysym}  
\usepackage{verbatim}

\usepackage{longtable}
\usepackage{float}
\restylefloat{table}	
\usepackage{accents}

\graphicspath{{./figures/}}    

\def\arraystretch{1.3}

\newif\iffinal
\finalfalse
\finaltrue

\newif\ifcomment
\commenttrue
\newif\ifpagebudget
\pagebudgettrue

\iffinal
	\commentfalse
	\pagebudgetfalse
\fi 

\newcolumntype{L}[1]%
{>{\raggedright\let\newline\\\arraybackslash\hspace{0pt}}m{#1}}
\newcolumntype{C}[1]%
{>{\centering\let\newline\\\arraybackslash\hspace{0pt}}m{#1}}
\newcolumntype{R}[1]%
{>{\raggedleft\let\newline\\\arraybackslash\hspace{0pt}}m{#1}}
\newcolumntype{T}[1]%
{>{\raggedright\let\newline\\\arraybackslash\hspace{0pt}}p{#1}}

%% file: abstract.tex
An accurate forecast of electric demand is essential for the optimal design of a generation system. For district installations, the projected lifespan may extend one or two decades. The reliance on a single-year forecast, combined with a fixed load growth rate, is the current industry standard, but does not support a multi-decade investment. Existing work on long-term forecasting focuses on annual growth rate and/or uses  time resolution that is coarser than hourly. To address the gap, we propose  multiple statistical forecast models, verified over as long as an 11-year horizon. Combining  demand data, weather data, and occupancy trends results  in a hybrid statistical model, i.e., generalized additive model (GAM) with a seasonal autoregressive integrated moving average (SARIMA) of the GAM residuals, a multiple linear regression (MLR) model, and a GAM with ARIMA errors model. We evaluate accuracy  based on: (i) annual growth rates of monthly peak loads; (ii) annual growth rates of overall energy consumption; (iii) preservation of daily, weekly, and month-to-month trends that occur within each year, known as the “seasonality” of the data; and, (iv) realistic representation of demand for a full range of weather and occupancy conditions.  For example, the models yield an 11-year forecast from a one-year training data set with a normalized root mean square error of 9.091\%, a six-year forecast from a one-year training data set with a normalized root mean square error of 8.949\%, and a one-year forecast from a 1.2-year training data set with a normalized root mean square error of 6.765\%.

%% file: intro.tex
\section{Introduction}
\label{intro}

\begin{figure}[htb!]
	\centering
    \includegraphics[scale=0.39]{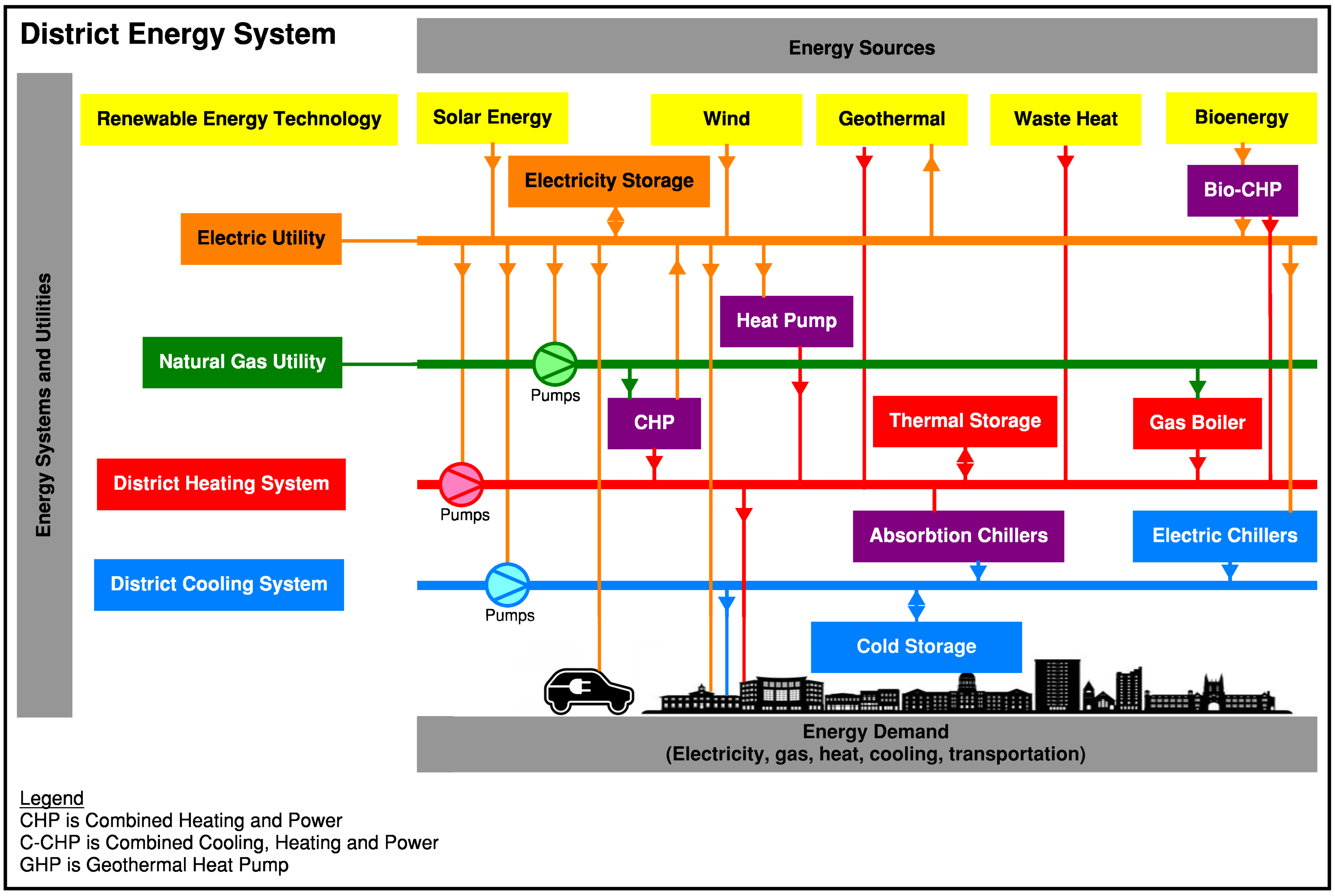}
    \caption{Possible interactions between energy networks, energy sources, and energy demands of a district energy system. The horizontal lines represent district systems. The vertical lines are the flow of electricity, natural gas, heating, and cooling to balance loads within the district. This figure is adapted from \citep{Wu}.} \label{fig:DES}
\end{figure}

\indent As the electrical grid across the U.S. approaches its distribution capacity and natural disasters that affect grid reliability become more prevalent, a focus toward resilience planning is emerging \citep{Bronec}. Owners are determining the value of resiliency in their systems and weighing the cost of distributed energy resource installations to protect their buildings and businesses from power outages \citep{NRELResilience}. A resilient power system is capable of islanding and operating independently from the grid \citep{NRELResilience}. A method to increase resiliency for meeting the demands of a building uses a centralized plant to supply  electricity, heating, and cooling to multiple buildings, thus creating a district energy system \citep{EESI}. Figure \ref{fig:DES} illustrates the possible interactions between renewable energy technologies, the electrical distribution network, the natural gas system, and the cumulative heating and cooling loads of all the buildings in a district energy system which is designed to facilitate opportunities to share energy production and storage resources. 

\begin{figure}[htb!]
	\centering
    \includegraphics[scale=0.6]{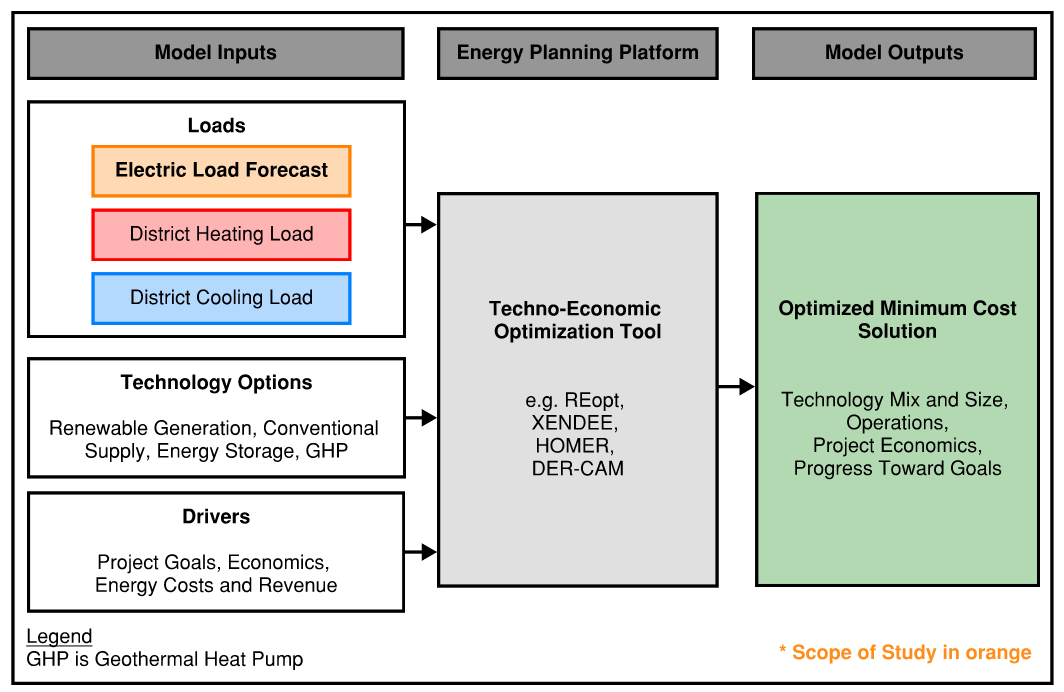}
    \caption{Flowchart of inputs and outputs of a renewable energy optimization model.} \label{fig:system}
\end{figure}

\indent Both the hourly electric load profile and the peak electric demand over the   lifetime of a district energy system influence the capacity needs of the energy production and storage resources. However, most district energy systems are designed with renewable energy optimization tools based on a single representative year of demand data \citep{Pecenak}, while the system may have a significant change in electric loads over its lifespan -- which can range between ten and 50 years. To accurately design the infrastructure and/or select optimal renewable energy technologies according to a standard process such as that given in Figure \ref{fig:system}, the electric demand forecast must capture future load growth. 

\indent While there are many data-driven models used to forecast building loads \citep{Bourdeau}, their application to forecasting electric demand in district energy systems is limited. Furthermore, few studies provide advice and general insights for selecting the appropriate statistical technique for long-range electric demand  forecasting using appropriate time fidelity. Hourly fidelity is crucial for modeling the electric loads during utility peak time-of-use windows because these determine electric demand charges and, correspondingly, impact energy technology system sizing. And, it is particularly important to have a long-term electric forecast for a district energy system  because of the longevity, the scale, and the potential  expansion of the system. Specifically, the sector of higher education institutions in the U.S. totals about 5 billion square feet of floor space with collective expenditures of approximately 6.5 billion per year on energy commodities, e.g., electricity and natural gas \citep{DOE}; also, many university campuses are hundreds of years old and tend to expand, rather than to be completely rebuilt. For example, the Colorado School of Mines has a detailed master plan of upcoming facility changes to align with state-of-the-art research needs, additional buildings to accommodate increased student population, and the plan for additional student housing; this master plan  provides guidelines for budgeting and fundraising with respect to capital improvements over the next ten years \citep{MinesMP}. Conversely, the master plan does not address how the expected increases in electric demand will impact the optimal sizing of renewable energy technology investments, which is the end use for this forecast framework. To address this shortcoming, a long-horizon electric demand forecast for use in a techno-economic optimization model allows investments to be informed by the demand of the current year, as well as by future demand. 

\indent This study develops a forecasting framework that automates the comparison and analysis of six statistical methods, as well as hybrid combinations of those methods, for modeling and forecasting electric demand using field data. In particular, we focus on developing an electric demand forecast model for implementation in renewable energy optimization software \citep{REopt} for grid-interactive district energy systems. The contributions of this paper are: (i) the development of a framework for  multi-year, long-horizon forecasting at hourly resolution for electric demand in a district energy system;  (ii) analysis and comparison of the electric demand forecast models; and, (iii) the application of this framework for the electric demand forecast to three case studies. The first case study is a small, public university with an annual energy use of 37,600 MWh and a peak hourly load of 6,900 kW. The case study is a representative district energy system with the following important features: a single-point electric utility connection; sustained year-after-year load growth; additional planned construction; and, an increased student population over the measured time horizon. The second case study is a large public university with an annual energy use of 211,000 MWh and a peak hourly load of 41,000 kW at the main campus electric meter. This case study has year-after-year load decrease; additional planned construction; and, an increased student population over the measured time horizon. The third case study is a medium-large, public university with an annual energy use of 135,000 MWh and a peak hourly load of 25,000 kW.

%% file: lit_review.tex
\section{Literature Review}
\label{lit_review}

\indent Electric demand forecasts are often system-specific and proprietary, excluding the existence of a universal forecast model; correspondingly, there is an absence of literature addressing this ever-growing topic. For district installations, the horizon may extend one or two decades to align with campus growth plans for constructing buildings. Bourdeau et al. \citep{Bourdeau} conduct an expansive review of available data-driven energy modeling and forecasting methods and find that fewer than 25\% of studies address a horizon of a year or longer. Of the long-term horizon studies, defined as a year or longer in this review, 53\%  use annual time-steps, 12\% use daily or weekly fidelity, 18\% are modeled at hourly fidelity, and 6\%  possess one-minute time-steps. Of the seven studies that consider horizon lengths of more than a year, all deploy annual time steps. The longest horizon of study with hourly fidelity is  one year, and, of these, only 30\% use an hourly or finer fidelity.   Zhu et al. \citep{Zhu} conduct a review of over 30 energy system forecast studies in which a month-ahead forecast is the longest   with hourly fidelity. 

\indent Physics-based building energy simulations are a common tool for electric, heating, and cooling demand estimations \citep{UBEM,Urban}. Rather than using overall energy modeling and uncertainty approaches to estimate the electric demand of each building within a district, data-driven models can  forecast  electric load demands \citep{Vaghefi}. Due to energy-performance and energy-monitoring requirements for most federally- or state-funded construction projects, such as government or college campuses, there is a significant amount of building-related data that are captured in  district energy systems \citep{GSA,CO}. 

\indent To compare the forecast methods of district system loads, we examine district heating and  cooling, in addition to district electricity demand forecasts, because they have similar factors and load patterns. Fang et al. \citep{Fang} forecast district heating loads and find that autocorrelation exists between observations as a function of the time lag between them. The forecast  with closest agreement to the data for their case study results from a linear regression model with a 168-hour (weekly) demand pattern \citep{Fang}. Daily, weekly, or annual trends  are important to incorporate into demand forecasts that are based on past data. While overall energy forecasting has been a focus of physics-based energy modeling, Bourdeau et al. \citep{Bourdeau} conduct an expansive review of available data-driven modeling methods and find that  heating and cooling loads are the primary focus of energy forecasting, and that energy end-uses such as electric demand are rarely considered. 

\indent Current platforms for renewable energy optimization lack a framework for modeling both electrical  and peak load growth for design horizons longer than one year \citep{Pecenak}. When developing a renewable energy or energy storage system that has a lifespan ranging from 10 to 30 years, the design horizon must, at a minimum, correspond to system longevity. Commonly, a single-year electric demand load profile is the primary input for available renewable energy optimization tools, such as REopt \citep{REopt}, XENDEE \citep{Pecenak}, HOMER Pro’s Optimizer \citep{HOMER}, and DER-CAM \citep{DER-CAM}. REopt uses only a single-year hourly load profile as an input with the ability to scale the load profile once to capture growth \citep{REopt}. XENDEE allows the user to input a percentage of load profile scaling for each year of the design horizon in a project; a different scale can be used for each year. The multi-year module in HOMER Pro allows the user to enter load growth in a simulation engine but not in an optimization model \citep{Pecenak}. The impact of multi-year load growth is not considered in the DER-CAM modeling software \citep{DER-CAM}. The platforms that attempt multi-year electric demand changes for distributed energy resource selection, such as XENDEE, do so by reducing the number of time steps in the optimization model to three representative days per year (weekday, weekend, and peak) or by providing an aggregate lifetime cost in lieu of an optimal renewable energy solution \citep{Pecenak}. However,  these approaches fail to fully capture the hourly demand changes over the design horizon. 

\indent Factors in district energy system load growth may include rising outdoor air temperatures, an increased number of buildings and occupants, electrification of natural gas equipment or cooling, and electric vehicle charging. In a sampling of regression-based overall building energy forecasts, common regressors consist of outdoor air temperature, outdoor relative humidity, solar radiation, time of day, day of the week, and previous power demand and energy consumption \citep{Bourdeau}. Less common factors include occupancy count and occupancy status \citep{Bourdeau}, both of which should be investigated for their effects on electric demand \citep{Vaghefi}.

\indent Data-driven methods are categorized into three branches: (i) statistical, (ii) artificial intelligence, and (iii) a hybrid \citep{Fathollahzadeh} based on which  inputs are used to model the behavior of the dependent variable: {\it exogenous variable based-models}, {\it past-and-present behavior models}, and {\it hybrid of exogenous variable and past-and-present behavior models}. Exogenous variable-based models  use linear or non-linear regression (e.g., multiple linear regression \citep{Apadula}) to represent the behavior of the target variable in relation to the exogenous variables \citep{Mavromatidis}. In a comprehensive review of published electric demand forecasting methods, Kuster et al. \citep{Kuster} find that exogenous variable-based models are most common for long-term forecasts. Regression models are efficient if the regressors are known and  carefully selected. However, studies usually neglect the complex interactions (e.g., multi-collinearity) between different exogenous variables. Past-and-present behavior models, e.g., seasonal autoregressive integrated moving average models (SARIMA), assume a linear form that imitates the behavior of the target variable based on the relationship between its present and past behavior as well as time lags \citep{Makridakis}. SARIMA models can analyze and forecast time series data with or without seasonality, but do not consider any exogenous variable’s effect in the modeling procedure. Hybrid-exogenous-variable and past-and-present  models \citep{Vaghefi} consider both the time series characteristic of the data as well as exogenous variables (e.g., linear regression with SARIMA errors). A hybrid approach has the benefit of including both regressor relationships and past-and-present data correlation, but it requires a significant number of estimated parameters (e.g., autoregressive terms). At the utility scale, Filik et al. \citep{Filik} formulate an hourly electric demand forecast for one year by nesting polynomial yearly, linear weekly, and linear hourly models, all of which mimic a hybrid model by using sinusoidal harmonic regressors in the weekly model.

\indent In addition to classical statistical methods, artificial intelligence methods could forecast electric demand \citep{Fathollahzadeh}. Lack of transparency in the relationship between input and output variables in an artificial neural network is the primary drawback \citep{Lee}. Kim et al. \citep{Kim} predict electricity consumption in a campus building using occupancy rates and weather conditions. They compare artificial neural networks and linear regression, and conclude that an artificial neural network can predict electricity consumption with higher accuracy when a small set of exogenous variables, or features, is incorporated into the model. Additional exogenous variables improve the accuracy of the linear regression model. A recurrent neural network consisting of long short-term memory cells successfully forecasts hourly regional utility electric demand for up to five years, but finds that the forecast had to be made on a rolling annual basis \citep{Agrawal}. This method has not been applied to longer forecasting scenarios or to district energy systems that show more variability. Another approach employs several types of artificial intelligence models, each chosen for their performance in hourly, daily, weekly, monthly, and annual time-steps and then layers the models to achieve forecast accuracy \citep{Wang}. However, artificial intelligence electric load forecast methods tend to compare the performance of proposed models with other artificial intelligence models rather than with classical statistical methods \citep{XIAO2024,CHEN2024,LI2024}. Additionally, hybrid combinations of artificial intelligence and statistical methods may improve the accuracy of statistical models or provide transparency to artificial intelligence methods, but they are not widely used due to their complex structure \citep{Fathollahzadeh}. Machine learning models can be constructed to incorporate interpretable steps, but there are still tradeoffs with the risk of undetected biases or undetected patterns within the data \citep{Bertsimas}.

%% file: methodology.tex
\section{Methodology}
\label{methodology}

\begin{figure}[ht]
	\centering
    \includegraphics[scale=0.69]{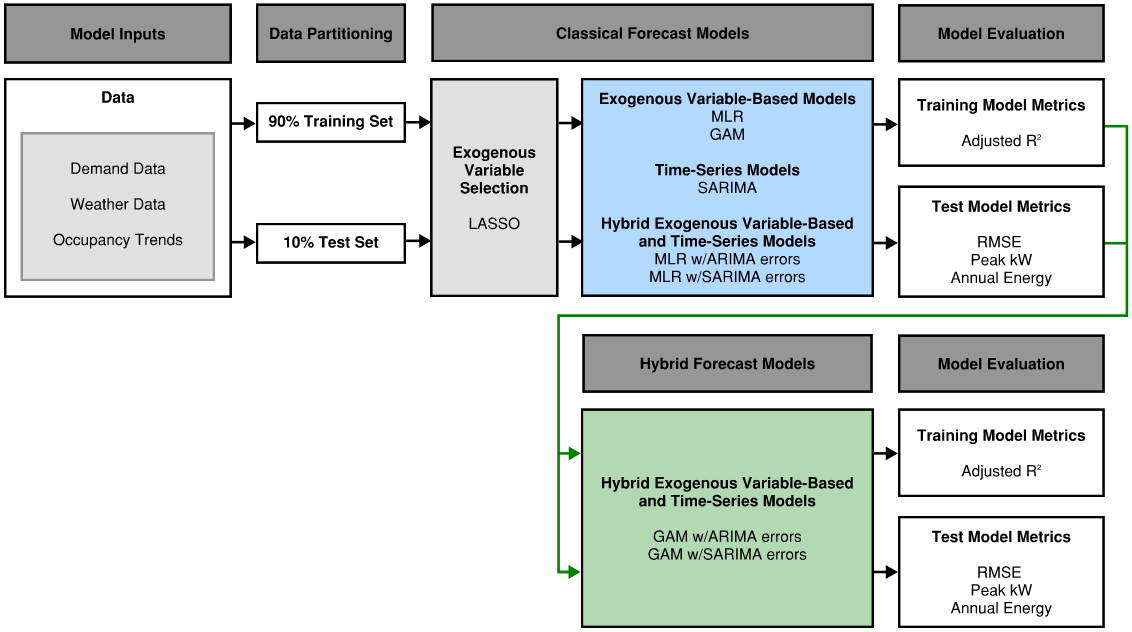}
    \caption{Framework flowchart for the development of an electric demand forecasting model and selection of the best model for implementation. Here, we include only our primary (and best performing) methodologies: the classical statistical techniques. The trained models include: least absolute shrinkage and selection operator (LASSO), multiple linear regression (MLR), generalized additive models (GAM), seasonal autoregressive integrated moving average (SARIMA), MLR with ARIMA errors, MLR with SARIMA errors, and hybrids of the aforementioned models.} \label{fig:flow}
\end{figure}

\indent Our forecasting framework describes and compares several data-driven statistical techniques used to model and forecast electric demand for a given district energy system. We compare both classical statistical and machine-learning model training techniques against test data; the model with the best fit, as determined by the quantitative metrics we outline below, is used to generate a long-range forecast. Figure \ref{fig:flow} shows our quantitative model selection corresponding to our primary methodology; the machine-learning models provide comparisons. We use electric demand, weather data, and occupancy trends for the district energy system. Based on the three categories of models, i.e., {\it exogenous-variable-based models}, {\it past-and-present-behavior models}, and {\it hybrid  exogenous-variable and past-and-present-behavior models}, we input the corresponding data (exogenous variables, past data, or both) into each model. The training set of input data is used for model development and the remainder is used for testing models. We use least absolute shrinkage and selection operator (LASSO) to analyze all potential   interactions between input (independent) variables and select the most significant, which are then used to construct exogenous-variable-based models. We run past-and-present-behavior models using the training set consisting of up to nine years for calibration and one year for evaluation to align with a commonly used 90\%-and-10\% data partition approach when the data set length allows; we also use a three-year test set (see Section \ref{subsec:comparative_metrics_simple}). Subsequently, we analyze combinations of models (hybrids) to capture the best characteristics of the classic models. Finally, we determine the  number of years required to develop an accurate forecast, where accuracy is given by the adjusted $R^2$ metric. 

\indent For a district energy system, the chosen forecast model must capture the following properties: (i) annual growth rates of monthly peak loads, because these determine electric demand charges; (ii) annual growth rates of overall energy consumption, because these determine energy and CO\textsubscript{2} charges; (iii) preservation of daily, weekly, and month-to-month (weather and load) trends that occur within each year, known as the “seasonality” of the data; and (iv) avoidance of oversmoothing. We evaluate test models using root mean square error (RMSE) for goodness of fit; we compare the forecast and actual data with respect to the peak electric demand values for coverage of the highest-energy-use days, and with respect to total energy use in a calendar year.

\subsection{Data inputs for electric demand}
\label{subsec:factors}

\indent We categorize the model inputs as follows: historical electric demand data represents the past behavior of the response variable; regressors are independent exogenous variables. The fidelity of the historic demand and exogenous variables matches the fidelity of utility time-of-use billing schedules, as well as the variability of renewable energy technology dispatch, to provide the required detail for optimal financial investments.

\subsubsection{Behavior of district energy system historical electric demand}
\label{subsubsec:historical_demand}

\indent The historical electric demand of the district energy system is a response variable that depends on several exogenous variables. Our exploratory analysis includes tests for normality, stationarity, skewness, and seasonality to provide insight into potentially viable statistical models.

\subsubsection{Exogenous Variables}
\label{subsec:exogenous}
\indent We categorize the exogenous variables for a model as follows: weather data, occupancy trends, and time-series seasonal and daily trends. This list of possible variables is based on past studies referenced in Section \ref{lit_review}, known correlations, and available data. 

\indent Those related to the weather data include:
\begin{itemize}
	\item {\it Temperature}: The actual meteorological year outdoor air temperature hourly data set for past years. Future outdoor air temperature is from statistically downscaled data from the Intergovernmental Panel on Climate Change General Circulation Models \citep{Chowdhury}.
    \item {\it Humidity}: The actual meteorological year outdoor relative humidity hourly data set for past years.	Future relative humidity is from statistically downscaled data from the Intergovernmental Panel on Climate Change General Circulation Models \citep{Chowdhury}.
\end{itemize}

\indent Those related to the occupancy trends include:
\begin{itemize}
	\item {\it Occupant Data}: This includes the hourly occupant schedule of the district energy system in terms of full-time equivalent students, staff, and faculty. A weight of 1.0 is assigned to the quantity of full-time students, staff, or faculty members, while 0.5 is used for half-time individuals.
	\item {\it Total Building Area}: This represents the total finished area of all buildings and can change over time. 
    \item {\it Energy Use Intensity}: As the total building area of a campus changes over time, so can the density of energy use per square foot. Buildings may have energy performance upgrades that decrease the value, or they may experience denser use  that increases the value over time. Energy use intensity is calculated by dividing the total energy consumed by the campus in one year (measured in kBTU) by the total gross floor area of the buildings on the campus (measured in square feet).
    \item {\it Day Category}: The categories of days for a university campus can be described as instruction on a weekday versus weekend, summer weekday versus weekend, and holiday weekday versus weekend. These day categories capture the behavior of thermostat setbacks, on-campus residents, and non-class-related events.	
    \item {\it Class Binary}: The behavior of occupants for a  university campus can be captured using a binary indicator for whether   or not there is instruction on any given day. This factor differentiates the days on which only staff are required to be on campus from the days with staff and students.
\end{itemize}

\indent We include exogenous variables to capture year-after-year trends and 24-hour daily trends   for the time-series behavior of  electric demand.
\begin{itemize}
	\item {\it Time-Step}:  The time-step exogenous variable  captures the linearly increasing trend in the model.

    \item {\it Cosine Term and Sine Term}: The following expressions for cosine and sine terms capture the seasonality of daily electrical load, where $\it{h}$ is the hour of the day and $\it t \in \mathcal{T}$ is the set of time steps:
    \begin{equation*}\label{none}
        Cosine Term = \cos \left( \frac{\text{2$\pi$$\it{h}$}}{\text{24}} \right) \forall t \in\mathcal{T}
    \end{equation*}
    \begin{equation*} 
        Sine Term = \sin \left( \frac{\text{2$\pi$$\it{h}$}}{\text{24}} \right) \forall t \in\mathcal{T}
    \end{equation*}
\end{itemize}

\indent These regressors are assessed for correlation with the electrical demand using regression selection techniques in the following subsection.

\subsection{Exogenous variable-based techniques}
\label{subsec:exog_models}
\indent This study analyzes a total of seven statistical models, divided into linear, non-linear and time-series forecasts. All of these are standard statistical methods documented in statistics literature; we choose them  for evaluation based on their suitability for this application.

\subsubsection{Linear regression models}
\label{subsubsec:LR_LASSO}
\indent Multiple linear regression (MLR) analysis is a statistical method to characterize the impact of selected exogenous variables (regressors) on a dependent variable \citep{Tibshirani}. There are no weights or penalties associated with linear regression model coefficients, and all exogenous variables are assigned an estimated coefficient to minimize the residual sum of squares, shown in equation \eqref{eq:RSS} \citep{Tibshirani}. Table \ref{tab:regression} introduces the general notation used in linear regression models, including a system-specific set of exogenous variables. The actual data is represented using $\it y_t$; coefficients $\beta_e$  are determined while creating the regression model.

\begin{longtable}[c]{llll}
\caption{Notation first used in equations \eqref{eq:LR}-\eqref{eq:GAM} and then in equations throughout the paper.} \label{tab:regression} \\
& \it Sets              &                                                               & Units  \\ \hline
& $\it t \in \mathcal{T}$    & The set of time steps; $\it t \in \{1,...\it T\}$              & [hour] \\
& $\it e \in \mathcal{E}$    & The set of exogenous variables; $\it e \in \{1,...\it E\}$    & [-]    \\
& & & \\

& \it Parameters            &                                                           & Units \\ \hline
& $\alpha$                  & intercept for linear regression model                     & [-]   \\
& $\beta_e$                 & coefficient of exogenous variable {\it{e}}                & [-]   \\
& & & \\

& \it Variables             &                                                           & Units \\ \hline
& $\it y_t$                 & target variable at time \it t                             & [W]   \\
& $\it \hat{y}_t$           & predicted target variable at time \it t                   & [W]   \\
& $\bar{y}$ & average value of variable across all $t \in \mathcal{T}$ & [W]   \\
& $\it x_{et}$              & value of exogenous variable {\it{e}} at time \it{t}       & [varies]     \\
& $\varepsilon_t$           & linear regression random error for target variable at time \it t  & [W]  \\
&&&\\
\end{longtable}

\begin{equation} \label{eq:LR}
    y_t = \alpha+\beta_1x_{1t}+\beta_2x_{2t}+\ldots+\beta_Ex_{Et}+\varepsilon_t \;\; \forall t\in\mathcal{T}
\end{equation}

\indent While MLR models minimize the sum of squared residuals for all exogenous variables, this method does not allow for the shrinkage toward zero of coefficients on less significant exogenous variables. Ridge regression and LASSO are the two most common regressor selection techniques with penalties which allow the coefficients for some regressors to approach zero \citep{Tibshirani}. In ridge regression, all exogenous variables are still assigned coefficients and remain in the model. The LASSO regressor selection technique allows the coefficients for some regressors to become exactly zero, thus reducing the cardinality of the set of regressors; this reduction improves model interpretability and reduces overfitting in high-dimensional settings \citep{Tibshirani, Bertsimas}.

\indent The LASSO model  penalizes the sum of the absolute values of the weights, which tends to reduce the weights of the least significant regressors to zero. LASSO minimizes the function given in equation \eqref{eq:LASSO} in which the first term is the same as the residual sum of squares; the second term is an $L _1$ norm added with the penalty of $\lambda$ for which, if the penalty is sufficiently large, some of the coefficients become zero. The appropriate value for the penalty is usually determined using a grid-search technique with a $k$-fold cross-validation \citep{Tibshirani}.  

\begin{equation} \label{eq:LASSO}
            \min \sum_{\it t \in \mathcal{T}} \left( y_t-\alpha-\sum_{\it e \in \mathcal{E}} \beta_e x_{et} \right)^2 + \lambda \sum_{\it e \in \mathcal{E}} |\beta_e |
\end{equation}

\subsubsection{Non-linear models}
\label{subsubsec:GAM}
\indent Generalized Additive Models (GAM)  estimate smooth, non-linear relationships between regressors and dependent variables \citep{Wood} and have been used in photovoltaic power prediction modeling  for short-term demand forecasts \citep{Sundararajan}. The approaches to modeling non-linear models include polynomial regression, step functions, regression splines, smoothing splines, and local regression \citep{Tibshirani}. For instances with multiple regressors, some or all of the regressors can be assigned using the aforementioned methods in lieu of linear coefficients and are represented as function $\it{f}$ in equation \eqref{eq:GAM}. The application for non-linear correlations between regressors and dependent variables can be identified through correlation studies.

\begin{equation} \label{eq:GAM}
    y_t = \alpha+f_1(x_{1t})+f_2(x_{2t})+\ldots+f_E(x_{Et})+\varepsilon_t  \;\; \forall t\in\mathcal T
\end{equation}

\indent Smoothing spline models produce forecasts in which the most recent observations carry a higher weight than older observations with an exponential decay. Exponential smoothing methods are commonly used for non-stationary data sets. A cubic regression spine takes both smoothness and local influence into consideration, and   penalizes deviations using the conventional integrated square second derivative cubic spline  \citep{Tibshirani}. We use these models in Section \ref{subsec:comparative_metrics_simple}.

\subsection{Past and present behavior based techniques: Time series forecast models}
\label{subsec:time_series}
\indent Besides linear and nonlinear exogenous variable-based models,   time series models can capture seasonal trends in the data. The following models use the relationship between past data and present data to forecast future data. Seasonal autoregressive (AR), integrated (I), moving average (MA) models (SARIMA) are a common category   \citep{Makridakis}. A SARIMA model has an order for each of the non-seasonal and seasonal AR, MA and differencing terms, as well as a season length in the format $(p,d,q)(P,D,Q)m$, as noted in Table \ref{tab:time_series}.

\indent The autoregressive order is the number of time steps in the past regression that are used to calculate the present value in the series. The moving average order is the number of time steps in the past forecast errors that are used to calculate the present value in the series. The integrated component refers to the existence of differencing in the model. The simple first differencing and the seasonal differencing both create stationarity by themselves. One or both of these  are expected to carry a first-order value in a SARIMA model to remove trends, seasonality, or both. The backshift operator shown in equation \eqref{eq:backshift_gen} is short-hand notation representing lagged time-series values. The equation for a generic SARIMA$(p,d,q)(P,D,Q)m$ model, with backshift notation, is shown in equation \eqref{eq:sarima}.

\begin{longtable}[c]{lll}
\caption{Notation first used in equations \eqref{eq:backshift_gen} and \eqref{eq:sarima} and then in equations throughout the paper.} \label{tab:time_series} \\
\multicolumn{2}{l}{\it Time Series Symbols} & Units  \\ \hline
  \it p                     & order of the non-seasonal AR lag polynomial           & [-] \\
 \it d                     & order of the non-seasonal differencing lag polynomial & [-] \\
 \it q                     & order of the non-seasonal MA lag polynomial           & [-] \\
 \it P                     & order of the seasonal AR lag polynomial               & [-] \\
 \it D                     & order of the seasonal differencing lag polynomial     & [-] \\
 \it Q                     & order of the seasonal MA lag polynomial               & [-] \\
 \it m                     & number of timesteps in a season                       & [-] \\
 $\phi_p$                  & non-seasonal autoregressive operator for lag polynomial \it p    & [-] \\  
 $\Phi_P$                  & seasonal autoregressive operator for lag polynomial \it P        & [-] \\
 $\theta_q$                & non-seasonal moving average operator for lag polynomial \it q    & [-] \\  
 $\Theta_Q$                & seasonal moving average operator for lag polynomial \it Q        & [-] \\  
 $\varepsilon_t$           & white noise value at time \it{t}                      & [kW] \\  
\end{longtable}

\begin{equation} \label{eq:backshift_gen}
        \text{Backshift Operator:} \;\; B^{m}y_t=y_{t - m}
\end{equation}
\begin{equation} \label{eq:sarima}
\begin{aligned}
   & (1-\phi_1B-...-\phi_pB^p)(1-\Phi_1B^{m}-...-\Phi_PB^{Pm})(1-B)^d(1-B^{m})^Dy_t = \\
   & (1+\theta_1B+...+\theta_qB^q)(1+\Theta_1B^{m}+...+\Theta_QB^{Qm})\varepsilon_t   \;\; \forall t\in\mathcal{T} \\
   \end{aligned}
\end{equation}

\indent A variation of the Hyndman-Khandakar algorithm for step-wise automatic ARIMA modeling \citep{Hyndman} is used to traverse the model space for the non-seasonal autoregression and moving-average orders. We use this automatic ARIMA technique for non-seasonal orders as a starting point for the SARIMA model orders. 

\indent We analyze the characteristics of data sets or the residuals of a training model using autocorrelation  and partial autocorrelation functions. There are several autocorrelation coefficients, corresponding to each panel in autocorrelation plots. For example, $r_1$ measures the relationship between $y_t$ and $y_{t-1}$, $r_2$ measures the relationship between $y_t$ and $y_{t-2}$, and so on. We review the autocorrelation functions of the best ARIMA model for the seasonal orders and season length to develop the best orders for seasonal components.


\subsection{Hybrid: Exogenous variables and past-and-present behavior-based techniques}
\label{subsec:hybrid}
\indent Hybrid models consider both the exogenous variables' relationships as well as the influence that a future value has on the past and present data. An MLR with a strong fit may have seasonality within the residuals,  indicating that a hybrid model may yield a forecast that more accurately captures the seasonality. A model called {\it linear regression with SARIMA errors} uses the coefficients of the MLR and then overlays the SARIMA model of the errors to capture all components of a forecast. The regression terms are affected by the autoregressive terms in this model. 

\indent Based on insights obtained from previous statistical techniques, SARIMAX uses significant  and available exogenous variables to address information remaining in the residuals of SARIMA. For SARIMAX, the same orders obtained with SARIMA are utilized and the exogenous variables from multiple linear regression are invoked. In this case, the regression terms are added to the SARIMA terms \citep{Hyndman}. While these example hybrid models are widely known, each hybrid model variant must be tailored to the specific application.

\subsection{Machine learning models}
\label{subsec:machine_learning}

For the purpose of comparison, machine learning methods may be used to develop a forecast model; the results of the model are analyzed with the same quantitative metrics as those with which the classical statistical techniques are. Recurrent Neural Networks (RNNs) are one of the most widely used models for performing time series predictions; however, they suffer from vanishing gradient descent, or a precipitous  decrease in the values of the gradient vector caused by the iterative multiplication of small numbers. A variation of RNNs, called Long Short-Term Memory (LSTM), is used to overcome this numerical instability by formulating long-term dependencies between training samples through activation functions \cite{Agrawal}.

\subsection{Evaluation of the forecasting performance}
\label{subsec:evaluating_model_performance}

\indent We compare the accuracy of different forecast models using an out-of-sample test, i.e., the data used in model ﬁtting (training) are different from those used in forecast evaluation (test). For our cases, we use up to nine years of hourly electric demand data to construct the initial forecasting models and one year of data for testing and validation in order to evaluate if the model captures the peaks and trends present in the data that would be absent with a smaller training data set. 

\indent The R\textsuperscript{2} metric is widely used for measuring regression performance, and reflects the portion of the variance that is explained by the regression \citep{Tibshirani}. However, as the number of regressors increases, the R\textsuperscript{2} metric increases accordingly, which can cause overfitting. Adjusted R\textsuperscript{2} is the  metric by which to judge  the best model without giving weight to the number of regressors used, and is therefore a better performance evaluation for a collection of models with several potential regressors. Adjusted R\textsuperscript{2}, using equations \eqref{eq:RSS} and \eqref{eq:adj_r_squared}, is the evaluation metric used for the quality of training model fit, where RSS is the residual sum of squares and TSS is the total sum of squares.

\begin{equation} \label{eq:RSS}
    RSS = \sum_{\it t \in \mathcal{T}} (y_t-\hat{y}_t)^2 , \;\;  TSS = \sum_{\it t \in \mathcal{T}} (y_t-\bar{y})^2 
\end{equation}

\begin{equation} \label{eq:adj_r_squared}
    R^2 = 1 - \left(\frac{\it RSS}{\it TSS}\right), \;\;  R^2_{adj} = 1 - (1 - R^2) \frac{\it \it T - 1}{\it \it T - \it E -1}
\end{equation}

\indent  We use training models to forecast the test year. We then compare the forecast and measured electric demand data for the last year using RMSE (see equation \eqref{eq:RMSE}), a common metric for examining the performance of the forecasting model; a smaller RMSE indicates better forecasting accuracy.

\begin{equation} \label{eq:RMSE}
    RMSE = \sqrt{ \frac{1}{T} \sum_{\it t \in \mathcal{T}}(y_t-\hat{y}_t)^2}
\end{equation}

\indent For  our application, we compare annual peak values (in kW), time-of-day of the peak  value (in kW), and annual energy use (in MWh) because they are important in the energy-cost application of this forecast.

%% file: results_test_case.tex
\section{Computational Statistical Results}
\label{sec:results_test_case}

\indent  Table \ref{tab:ThreeCases} shows the  three case studies used to demonstrate our framework; all possess district energy systems from three different climatic zones with energy use per person ranging from 5.4 to 6.2 MWh: the  Colorado School of Mines (Mines), the University of California Davis (UCD), and Clemson University (Clemson). The Mines primary electrical meter data is for the years between 2008 and 2019 while that from UCD is from 2017 through 2023 and that from Clemson is from November 2021 through March 2024.

\begin{table}[H]
    \caption {Comparison of the three case study locations and characteristics}\label{tab:ThreeCases} 
    \centering 
    \begin{tabular}{l c c c c c c c}
        \hline  
        School         & Dataset& Campus    & Student     & Climate Zone  & Peak        & Energy   & Energy Use     \\ 
        (last year     & Length & Area      & Population  & \& Humidity   & Demand$^{\triangle}$ & Use$^{\triangle}$  & per Person$^{\triangle}$     \\ 
        of dataset)       & (years)& (acres)   & (people)    & (IECC)$^{\dagger}$& (kW)    & (MWh)  & (MWh/person)   \\ \hline
        Mines (2019)   & 12     & 500       & 6,287       & 5B            & 6,989       & 39,059   & 6.213          \\   
        UCD (2023)     & 7      & 5,300     & 39,679      & 3B            & 42,811      & 212,946  & 5.367          \\          
        Clemson (2023) & 2.4    & 1,400     & 22,566      & 3A            & 25,072      & 135,321  & 5.997          \\ \hline   
    \end{tabular}

     $^{\dagger}$The IECC Climate Zones range from zero to eight with zero being the warmest. The IECC Moisture Regimes are moist (A), dry (B), and marine (C). $^{\triangle}$The values in this column correspond to the last year in each dataset.

\end{table}

\indent The Mines campus, located in Golden, Colorado, USA uses a district energy system with electric, gas, chilled water, and steam loops serving the buildings \citep{MinesMP}. The system possesses the following important features: a single-point electric utility connection, sustained year-after-year load growth, and an increasing student population over the measured time horizon. The primary distribution loop serves the electrical demand of 52 buildings at the time of the data measurement, which was conducted at 15-minute fidelity \citep{MinesData}. Data of this fidelity and spanning this length of time is rare.  

\indent The UCD campus, located in Davis, California, USA uses a district energy system for the main academic campus with electric, gas, chilled water, central steam, a solar farm, and rooftop solar power. The electrical demand data is measured at an hourly fidelity and the solar production is added back to the demand to acquire the total electric demand. A unique attribute of this data is that the peak demand decreases year-after-year despite the increase in full-time-equivalent occupancy \citep{UCD}.

\indent The Clemson University campus, located in Clemson, South Carolina, USA uses a district energy system with a 15MW combined-heat-and-power steam co-generation plant, central chilled water, electricity, and natural gas. The electric energy demand data measurement is conducted at 15-minute fidelity \citep{Clemson}.

For all three cases, the campus full-time-equivalent faculty and student data are extracted from annual audits. We draw class schedules from historic academic calendars to determine occupancy with respect to time. Supplemental detailed analysis and framework results for all three cases are described in \ref{appendixB}; we provide additional detailed data analysis for Mines, comparisons between model categories for UCD, and qualitative forecast comparisons for Clemson. All model runs employ a computer with a 64-bit operating system, Core i7-12800 (2.4 GHz) processor, and 32 GB of RAM. Statistical analysis was carried out using CRAN \citep{teamR} in R studio version 2023.06.1 \citep{Rstudio}.

\subsection{Electrical demand characteristics}
\label{subsec:electrical_mines}

\begin{figure}[htb!]
	\centering
    \includegraphics[scale=0.55]{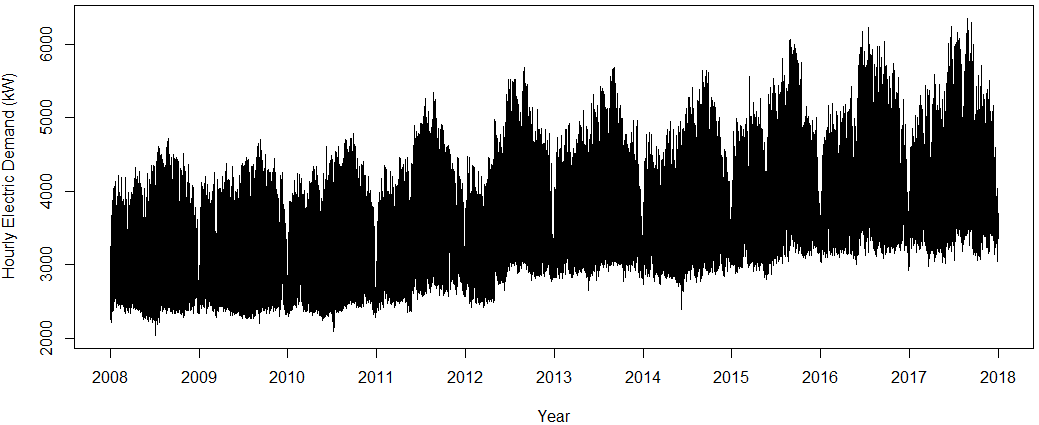}
    \caption{Hourly electrical meter data measured at Mines' primary meter.} \label{fig:CampusLoad}
\end{figure}

\indent We aggregate the primary electric meter demand data from the Mines and Clemson campuses from the original fidelity of 15-minute intervals, into hours by taking the sum of the four 15-minute intervals for each hour. The UCD primary electric meter demand data possesses hourly fidelity. We check the hourly campus meter data for zeros and NAs before log-transforming it. We replace zero values, which would correspond either to a power or metering outage, with the mean of the hour before and the hour after the value. In the event of an outage lasting over one hour, we replace the zero values with the scaled hourly data from the previous day; the demand values before and after the outage compared to the reference day determine the scaling factor. Figure \ref{fig:CampusLoad} shows the hourly aggregated Mines meter data at the primary Xcel Energy utility meter.
 
\begin{figure}[htb!]
	\centering
    \includegraphics[scale=0.49]{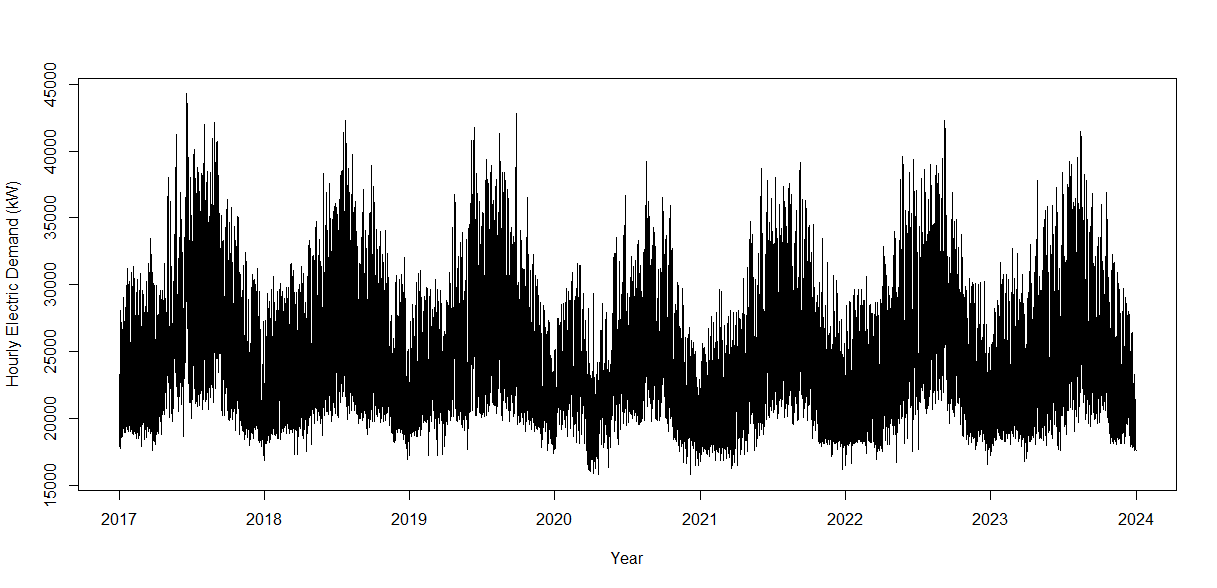}
    \caption{Hourly electrical meter data measured at University of California Davis' primary meter. } \label{fig:UCDLoad}
\end{figure}

\indent Figure \ref{fig:UCDLoad} shows the electric demand data for UCD. The amount of energy used per square foot of the buildings connected to the primary electric meter decreases over time due to concerted energy efficiency improvement efforts.

\begin{figure}[htb!]
	\centering
    \includegraphics[scale=0.52]{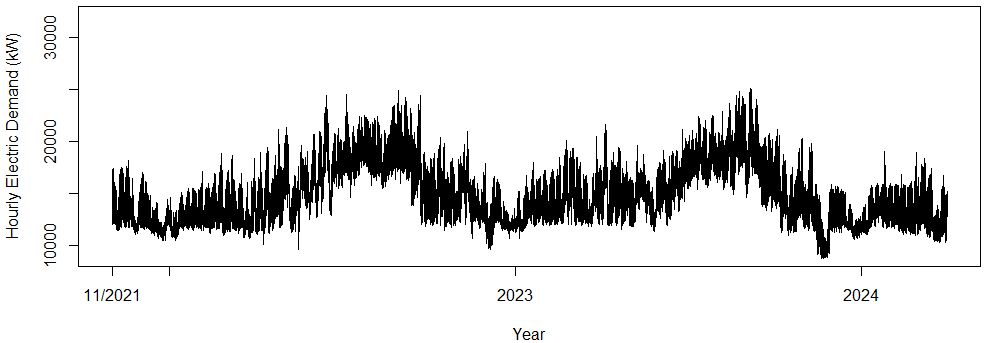}
    \caption{ Hourly electrical meter data measured at Clemson's primary meter.} \label{fig:ClemsonLoad}
\end{figure}

\indent Figure \ref{fig:ClemsonLoad} depicts the third data set used to validate the framework, from Clemson. The measured data set has a length of 2.4 years; the forecast framework was applied to a 1.2-year training set (through end of 2022) and a one-year test set (for the year 2023).  The test set is truncated to one year because the energy metrics and comparisons apply to only full years.

\indent For all three cases, we check the metered data for normality using the Kolmogorov-Smirnov test; the p-values of the tests are much smaller than 0.05, which indicates that the data is non-normally distributed. To reduce the fluctuation of the data values and the skewness and to exaggerate the periodic structure, we perform a logarithmic transformation for each case.

\newpage
\subsection{Selected exogenous variables}
\label{subsec:MinesExog}

\indent The {\it Occupancy} is an estimated hourly occupancy of all of the buildings on each campus. This encompasses the historical full-time-equivalent student, faculty, and staff information available for each school year, as well as the number of students living on campus. The hourly actual meteorological year outdoor air {\it Temperature} is collected for each case study location. 

\indent To develop the most accurate model for the electricity load demand, we must add  significant exogenous factors other than occupancy and temperature. For example, the occupancy rates are estimated, but whether class is in session or it is a weekend day are determined by the academic calendar. Adding categorical variables provides an explanation for days with similar outdoor air temperatures that might have different electric demands. 

\indent We define the additional initial exogenous variables as follows: {\it Humidity} is the actual meteorological year outdoor relative humidity hourly data set for the city associated with each case study. {\it Total Building Area} is the total building area of the campus as buildings are connected to the primary electric meter over time. {\it Energy Use Intensity} (EUI) is the amount of energy used per square foot annually. {\it Day Category} is defined as the following six categories: spring and fall semester weekday, spring and fall semester weekend, summer weekday, summer weekend, holiday break weekday, and holiday break weekend. These day categories capture the behavior of thermostat setbacks, on-campus residents, and non-class-related events. {\it Class Binary} assumes a value of one when it is a standard class day, and zero for all other days. There are many days during the year on which most faculty, staff, and administration are on campus but classes are not held. 

\indent This study uses LASSO regression to select the appropriate exogenous variables. LASSO is a modification of linear regression that minimizes the complexity of the model by limiting the sum of absolute values of the model coefficients \citep{Deng}. We perform k-fold cross-validation to find the optimal $\lambda$ value (see Equation \ref{eq:LASSO}) to minimize the mean-square error. Then, using the best $\lambda$, we determine the LASSO regression coefficients.

\begin{longtable}[c]{l | c c c c c c c c c c}
\caption{\label{tab:Cases} Comparison of the selected exogenous variables for the three case studies. The selected variables are marked with an 'X.' Note: EUI = energy use intensity.} \\
\multicolumn{1}{l}{} &
\multicolumn{10}{c}{Exogenous Variables} \\  \hline\hline
                    & Time & Occup- & Outdoor  & Relative   & Day      & Class   & Building &     & Cosine & Sine  \\ 
        Case        & Step & ancy   & Temp.    & Humidity   & Category & Binary  & Area     & EUI & Term   & Term  \\ \hline
        Mines       & X    & X      & X        &            & X        & X       &          &     & X      & X     \\ 
        UCD         & X    & X      & X        &            & X        & X       &          &  X  & X      & X     \\          
        Clemson     & X    & X      & X        & X          & X        & X       &          &     & X      & X     \\ \hline  
\end{longtable}

\indent For Mines, the outdoor relative humidity, total building area, and energy use intensity return a coefficient of zero in the LASSO regression; hence, we omit them from the model. While the correlations are close in value, the scatterplot shows a non-linear correlation between electric load demand and outside temperature. The increase in electric demand with higher temperatures is due to Mines using electricity for cooling, but natural gas for heating.

\indent For UCD, in addition to the exogenous variables {\it{Time-Step}, Occupancy, Temperature, Day Category, Class Binary, Cosine Term,} and {\it Sine Term} selected for Mines, the {\it{Energy Use Intensity}} remains after the LASSO selection is performed. The correlations  are linear  for all of the exogenous variables except temperature. The {\it{Energy Use Intensity}} exogenous variable for this campus captures the energy efficiency measures implemented by the university. 

\indent The Clemson LASSO selection retains exogenous variables {\it{Time-Step}, Occupancy, Temperature, Day Category, Class Binary, Cosine Term, Sine Term}, and {\it{Humidity}} after the LASSO selection is performed.  Again, the correlations  are linear  for all of the exogenous variables except temperature.  The exogenous variable outdoor humidity has significance due to the electric demand associated with a warm, humid climate.

\subsection{Comparative metrics of exogenous variable regression models}
\label{subsec:comparative_metrics_simple}

\indent An MLR is applied to the remaining exogenous variables after the LASSO selection for the respective cases. To further improve the fit of the linear regression, we include non-linear coefficients for exogenous variables with non-linear correlation in a GAM model. We train a matrix of combinations of spline and spline cubic regression functions on the exogenous variables. The best GAM model fit configurations are with GAM\textsubscript1, which has a spline cubic regression on the temperature regressor, and GAM\textsubscript2, which has a spline cubic regression on both the temperature and occupancy regressors.

\indent Table \ref{metrics1} compares the accuracy of each analyzed simple technique to model and forecast electric demand for the nine-year training set for Mines, a six-year training set for UCD, and a 1.2-year training set for Clemson. We use a one-year test data set for all three cases. For the Mines training data set, GAM\textsubscript1 has the highest adjusted-$R^2$ value. And, while GAM\textsubscript1 has the best RMSE values on the one-year test data set, GAM\textsubscript2 has the best RMSE on the three-year test data set. The residuals indicate 24-hour trends that are not captured with the model. The results are similar for the residuals of all three simple models. The GAM\textsubscript1 has the lowest NRMSE and closest annual energy use and GAM\textsubscript2 has the closest peak value, so we examine both models further in the following section.

\indent For the UCD data set, GAM\textsubscript2 has the highest training adjusted-$R^2$ value. The GAM\textsubscript1 model has a peak value closer to the actual value than the GAM\textsubscript2 model, but the GAM\textsubscript2 model has the lowest error and closest annual energy use values. The Clemson data set has a similar combination of best training and test metrics between the GAM\textsubscript1 and GAM\textsubscript2 models. While GAM\textsubscript2 has the highest  training adjusted R\textsuperscript{2}, GAM\textsubscript1 has the lowest test set error. 

\begin{longtable}[c]{ll | cc | ccc }
\caption{Comparative metrics of training sets of varying lengths and one-year test sets of simple exogenous variable-based models. The GAM\textsubscript1 model has a spline cubic regression on the temperature regressor and the GAM\textsubscript2 model has a spline cubic regression on both the temperature and occupancy regressors. Test RMSE values are normalized by the mean of each data set, yielding the normalized root mean squared error (NRMSE). The peak values are a percentage of the actual peak kW. The energy use values are a percentage of the actual annual energy use. Bold indicates the best results.} \label{metrics1}\\
\multicolumn{2}{r}{} &
\multicolumn{2}{c}{Training set} &
\multicolumn{3}{c}{One-year test set} \\ \hline
        &                    &              &                 & NRMSE            & Peak              & Energy Use       \\ 
Case    & Model              & Set length   & Adj-$R^2$       & (\%)             & (\%)              & (\%)             \\ \hline
Mines   & MLR                & Nine years   & 0.7908          & 3.208            & 96.38             & 103.2            \\
        & GAM\textsubscript1 &              & \textbf{0.8830} & \textbf{2.946}   & 107.7             & \textbf{102.8}   \\
        & GAM\textsubscript2 &              & 0.8682          & 3.393            & \textbf{103.4}    & 103.4            \\ \hline
UCD     & MLR                & Six years    & 0.6033          & 1.914            & 90.83             & 102.8            \\
        & GAM\textsubscript1 &              & 0.6099          & 2.063            & \textbf{102.6}    & 103.0            \\
        & GAM\textsubscript2 &              & \textbf{0.6323} & \textbf{1.853}   & 103.4             & \textbf{102.7}   \\ \hline
Clemson & MLR                & 1.2 years     & 0.7260          & 6.998            & 93.24             & 142.8            \\
        & GAM\textsubscript1 &              & 0.8150          & \textbf{5.363}   & 105.2             & \textbf{142.1}   \\
        & GAM\textsubscript2 &              & \textbf{0.8230} & 5.558            & \textbf{104.4}    & 143.6            \\ \hline

\end{longtable}

\subsection{Comparative metrics of time series models}
\label{subsec:comparative_metrics_time-series}
\indent Linear and non-linear regression models that use only exogenous variables to forecast electric demand do not adequately capture the seasonal behavior of the actual campus meter data; therefore, we also evaluate time series models. Our analysis includes ARIMA and SARIMA, but alone, they do not capture the correlation with the exogenous variables, and, thus, we do not portray the results here. Table \ref{metrics2} compares the accuracy of a sample of analyzed non-seasonal and seasonal time-series models and forecast electric demand for the training and test data sets for all three cases. These models have flat behavior, failing to capture the peak values of the actual meter data. With these shortcomings of each classical statistical forecast model used independently, we evaluate the hybrid models in the following section. 

\begin{longtable}[c]{ll | cc | ccc }
\caption{Comparative metrics of training sets of varying lengths and one-year test sets of time series models. Test RMSE values are normalized by the mean of each data set, yielding the normalized root mean squared error (NRMSE). The peak values are a percentage of the actual peak kW. The energy use values are a percentage of the actual annual energy use. }\label{metrics2}\\
\multicolumn{2}{r}{} &
\multicolumn{2}{c}{Training set} &
\multicolumn{3}{c}{One-year test set} \\ \hline
        &                         &            &            & NRMSE     & Peak       & Energy Use \\ 
Case    & Model                   & Set length & Adj-$R^2$  & (\%)      & (\%)       & (\%)       \\ \hline
Mines   & ARIMA(0,1,5)            & Nine years & 0.9669     & 30.21     & 48.94      & 74.67      \\
        & ARIMA(2,1,1)            &            & 0.9660     & 31.74     & 48.42      & 72.86      \\
        & SARIMA(0,1,5)(0,0,1)24  &            & 0.9781     & 30.02     & 50.40      & 74.90      \\
        & SARIMA(2,1,1)(0,0,1)24  &            & 0.9778     & 31.36     & 49.72      & 73.33      \\ \hline
UCD     & ARIMA(4,1,0)            & Six years  & 0.9138     & 34.20     & 41.67      & 71.34      \\
        & ARIMA(2,1,1)            &            & 0.9138     & 34.13     & 41.68      & 71.43      \\
        & SARIMA(0,0,0)(1,1,0)24  &            & 0.6154     & 23.36     & 56.44      & 85.62      \\
        & SARIMA(0,1,5)(0,0,1)24  &            & 0.9235     & 33.04     & 41.47      & 72.73      \\ \hline
Clemson & ARIMA(5,1,1)            & 1.2 years   & 0.9817     & 23.44     & 49.15      & 76.56      \\
        & ARIMA(2,1,1)            &            & 0.9778     & 32.62     & 44.53      & 67.32      \\
        & SARIMA(5,1,1)(1,0,1)24  &            & 0.9862     & 48.35     & 49.33      & 51.65      \\
        & SARIMA(2,1,1)(1,0,0)24  &            & 0.9809     & 31.97     & 45.50      & 68.04      \\ \hline
\end{longtable}

\begin{figure}[htp]
	\centering
        \captionsetup{justification=centering}
	\subfloat[Mines measured campus electrical demand data and a sample of the first 10 days of the following training models: MLR, GAM\textsubscript1, GAM\textsubscript2, and SARIMA with the associated adjusted $R^2$.]{\includegraphics[scale=.5]{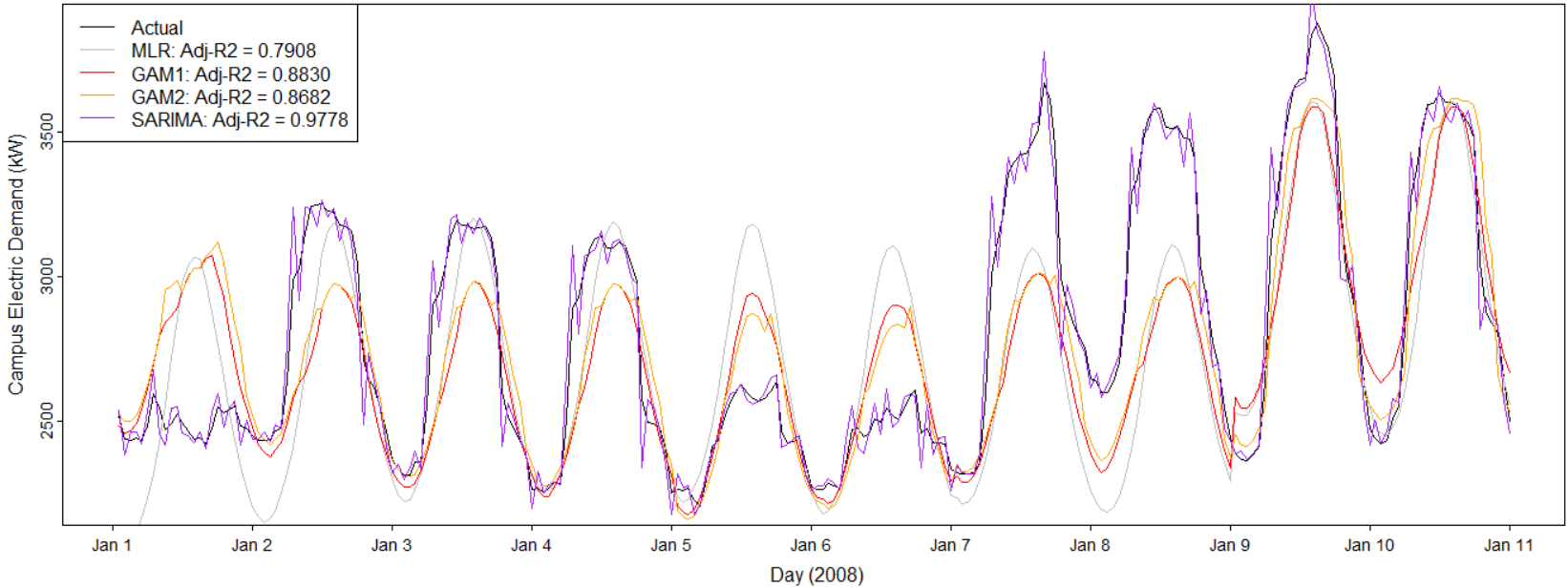} \label{sub:TrainMetric}} \\
    \subfloat[Mines measured campus electrical demand data and a sample of the first 10 days of the following test models: MLR, GAM\textsubscript1, GAM\textsubscript2, and SARIMA with the associated adjusted $R^2$.]{\includegraphics[scale=.5]{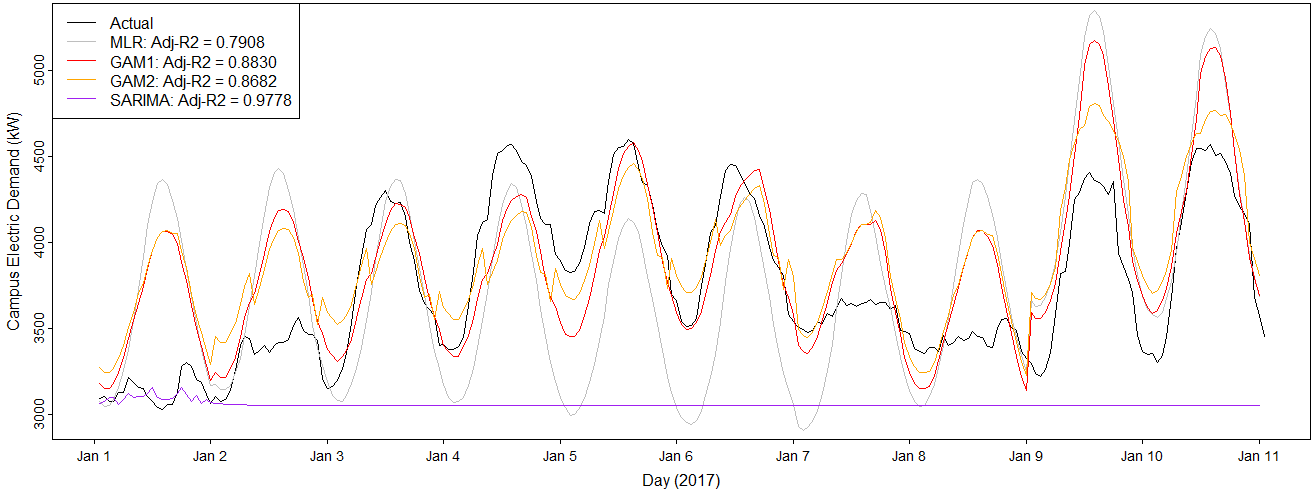} \label{sub:TestMetric}}\\   
	\caption{Actual and calculated campus electric energy usage for the a) training and b) test models using Mines data for the first 10 days of the training and test sets to analyze time-series model behavior. Figure a) shows January 1 through January 10, 2008. Figure b) shows January 1 through January 10, 2017.} 
	\label{fig:TrainTestComp}
\end{figure}

\indent The adjusted coefficient of determination (adjusted R\textsuperscript{2}) metrics for the time series models for all three cases  indicate that the training models explain more that 90\% of the total variability within the training data. Figure \ref{sub:TrainMetric} shows the first 10 days of a sample of training models for the Mines case. While this metric indicates the quality of the training fit, it does not always correspond to the quality of the test fit. Figure \ref{sub:TestMetric} shows the first 10 days of the test models associated with the training models. Note that the time-series SARIMA model, while having the highest adjusted R\textsuperscript{2} of the models, loses all seasonality and trends in the test set and the forecast converges to a constant value after around 24 time steps.

\newpage

\subsection{Comparative metrics of hybrid models}
\label{subsec:comparative_metrics_hybrid}
 
\indent Hybrid models can capture daily trends more accurately than simple regression models. Table \ref{metrics3} compares the accuracy of each analyzed hybrid technique to model and forecast electric demand for an appropriate-length training and one-year test data sets. Several variations of the hybrid models presented in Figure \ref{fig:flow} are run for each case study; from these, we tabulate four models for each campus location with the best metrics. 

For the Mines training data set, the MLR+ARIMA model has the highest adjusted $R^2$ value. For the one-year test data sets, GAM\textsubscript1+SARIMA has lower RMSE values than MLR+ARIMA and all other models. The GAM\textsubscript1+SARIMA model captures and exceeds the peak values and closely aligns with the total energy use.  As a metric of evaluation, the time-of-day of the peak values for all models is 1:00pm, which aligns with the actual meter data (see Section \ref{subsec:evaluating_model_performance}). Overall, GAM\textsubscript1+SARIMA provides the most accurate results among all analyzed techniques in longer test data sets. The formulation for the GAM\textsubscript1+SARIMA model is provided in \ref{appendix}.

\begin{longtable}[c]{ll | cc | ccc }
\caption{Comparative metrics of training sets of varying lengths and one-year test sets of hybrid statistical models. Test RMSE values are normalized by the mean of each data set, yielding the normalized root mean squared error (NRMSE). The peak values are a percentage of the actual peak kW. The energy use values are a percentage of the actual annual energy use. Bold indicates the best results.} \label{metrics3}\\
\multicolumn{2}{r}{} &
\multicolumn{2}{c}{Training set} &
\multicolumn{3}{c}{One-year test set} \\ \hline
        &                                           &               &                   & NRMSE         & Peak          & Energy Use     \\ 
Case    & Model                                     & Set length    & Adj-$R^2$         & (\%)          & (\%)          & (\%)           \\ \hline
Mines   & MLR+ARIMA(0,0,1)                          & Nine years    & \textbf{0.9746}   & 3.184         & 125.4         & 124.0          \\
        & MLR+SARIMA(0,0,1)(1,0,0)24                &               & 0.9495            & 3.371         & \textbf{97.94}& 103.4          \\
        & GAM\textsubscript1+SARIMA(5,1,1)(1,0,0)24 &               & 0.8460            & \textbf{0.4349}& 104.1        & \textbf{99.56} \\
        & GAM\textsubscript2+SARIMA(0,1,5)(1,0,0)24 &               & 0.8720            & 17.61         & 82.18         & 82.40          \\ \hline
UCD     & MLR+SARIMA(1,0,0)(1,0,0)24                & Six years     & \textbf{0.9351}   & 1.884         & 77.84         & 102.8          \\
        & GAM\textsubscript2+ARIMA(4,1,2)           &               & 0.6963            & 1.253         & \textbf{100.3}& \textbf{99.62} \\ 
        & GAM\textsubscript1+SARIMA(5,1,2)(1,0,0)24 &               & 0.6751            & 1.223         & 101.9         & 102.1          \\ 
        & GAM\textsubscript2+SARIMA(4,1,0)(1,0,0)24 &               & 0.6371            &\textbf{0.4319}& 102.0         & 101.3          \\ \hline
Clemson & MLR+ARIMA(5,1,0)                          & 1.2 years      & \textbf{0.9829}   & 14.05         & 64.71         & 85.91          \\
        & GAM\textsubscript1+ARIMA(5,1,1)           &               & 0.8297            & 7.891         & 91.92         & 92.10          \\ 
        & GAM\textsubscript2+ARIMA(5,1,2)           &               & 0.8335            & \textbf{6.765}& \textbf{92.20}& \textbf{93.23} \\ 
        & GAM\textsubscript2+SARIMA(5,1,2)(0,0,1)24 &               & 0.8412            & 9.743         & 87.39         & 90.25          \\ \hline
\end{longtable}

\begin{figure}[htb!]
	\centering
        \captionsetup{justification=centering}
	\subfloat[GAM\textsubscript1+SARIMA: Year: 2017]{\includegraphics[scale= 0.45]{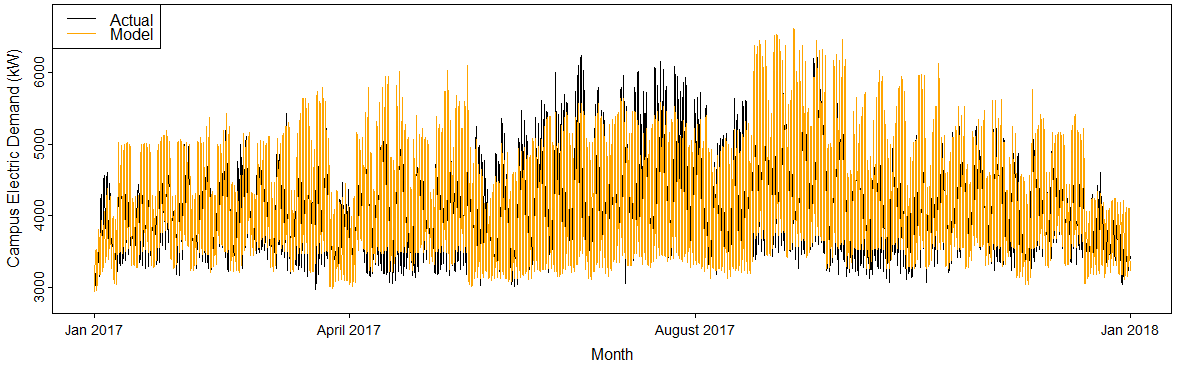} \label{sub:GAMSARyear}} \\
    \subfloat[GAM\textsubscript1+SARIMA: Month of August]{\includegraphics[scale= 0.45]{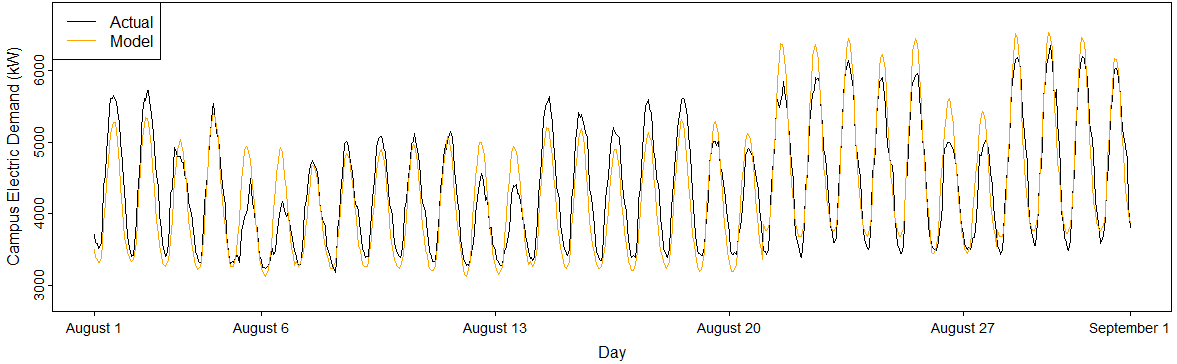} \label{sub:GAMSARmonth}}\\ 
	\subfloat[GAM\textsubscript1+SARIMA: Week of August 13th]{\includegraphics[scale= 0.45]{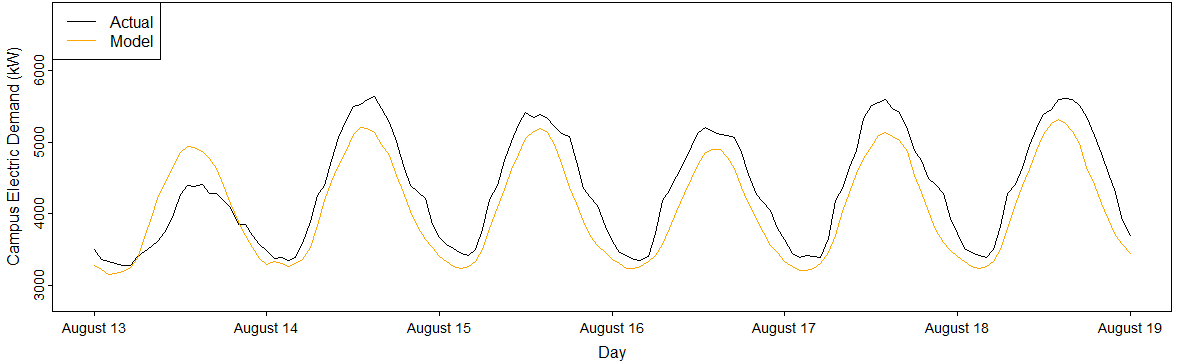} \label{sub:GAMSARweek}}\\
    \subfloat[GAM\textsubscript1+SARIMA: Day of August 16th]{\includegraphics[scale= 0.45]{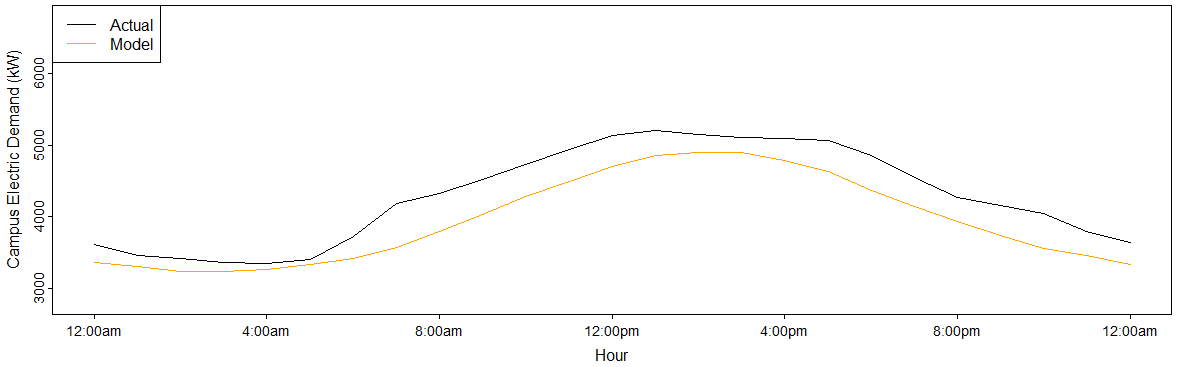} \label{sub:GAMSARday}}\\
	\caption{Mines predicted versus actual electric demand of (a) one year, (b) one month, (c) one week, and (d) one day. The representative month of August has the highest electric demand, reflects days with the highest outdoor temperature and occupant activity, and possesses historically peak electric demand.}
	\label{fig:GAMSARIMA}
\end{figure}

\indent The Mines hybrid model in Figure \ref{fig:GAMSARIMA} shows the forecast results for the GAM\textsubscript1+SARIMA model for a one-year test dataset, specifically the model alignment with actual data in representative monthly, weekly, and daily load demand patterns. The month of August historically has the highest electric demand values and represents the essential load for optimal sizing of renewable energy technologies. In addition, for most district energy systems, including at Mines, the utility billing demand charges are based on the highest demand in any given month on a non-holiday weekday between 2:00pm and 6:00pm. If the highest forecast demand value in each month is lower than the actual value, models that evaluate the cost savings of a set of renewable energy technologies will provide a conservative estimate.  

\indent The one-year NRMSE for the GAM\textsubscript1+SARIMA model shows an improvement in the forecast accuracy and in the total energy use estimate over the GAM\textsubscript1 model given in Section \ref{subsec:comparative_metrics_simple}. 

\newpage 

\begin{figure}[htb!]
	\centering
    \includegraphics[scale=0.42]{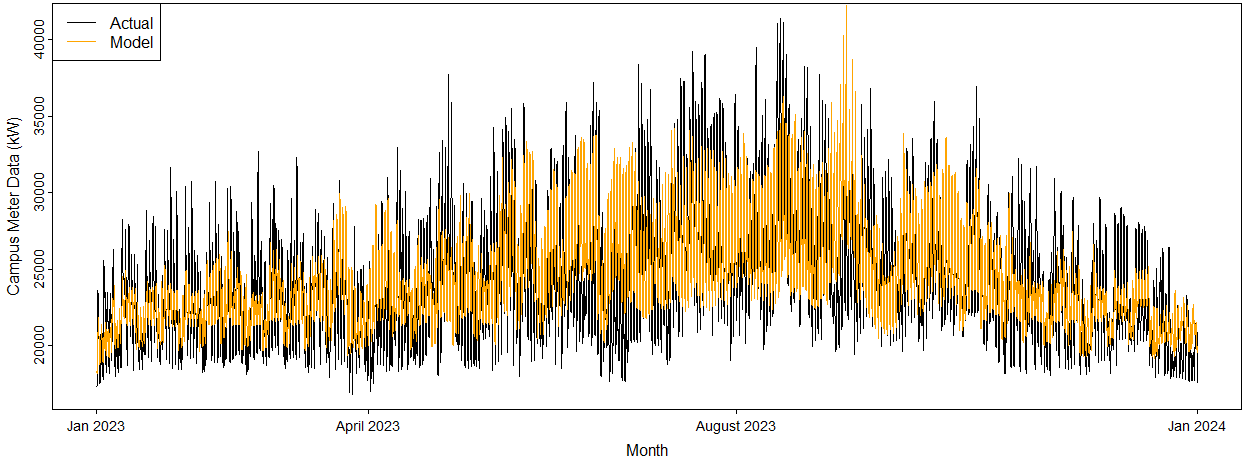}
    \caption{UCD GAM\textsubscript2+SARIMA one-year test. } \label{fig:UCDGAM2SAR}
\end{figure}

\indent Figure \ref{fig:UCDGAM2SAR} shows the UCD hybrid forecast model GAM\textsubscript2+SARIMA. While other hybrid models have slightly closer peak and annual energy use metrics, GAM\textsubscript2+SARIMA has the lowest NMRSE.

\begin{figure}[htb!]
	\centering
    \includegraphics[scale=0.42]{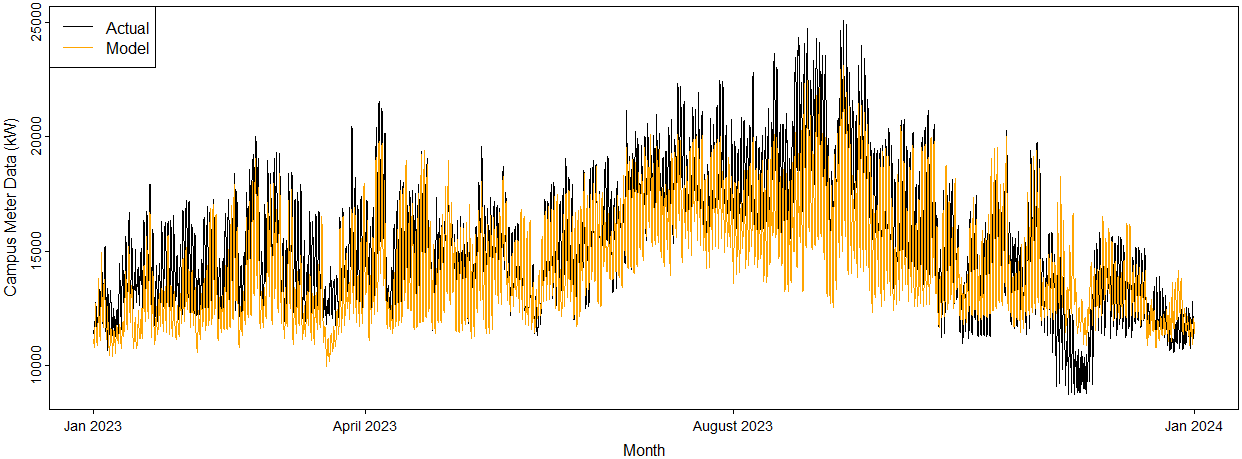}
    \caption{Clemson GAM\textsubscript2+ARIMA one-year test. } \label{fig:ClemsonGAM1}
\end{figure}

\indent The Clemson hybrid model with the best performance is shown in Figure \ref{fig:ClemsonGAM1}. This GAM\textsubscript2+ARIMA model has the lowest NRMSE and the closest peak and annual energy use values to the actual data. 

\subsection{Long-horizon electric demand forecast}
\label{subsec:long_horizon_forecast}

\indent From the model evaluation framework results summarized in Sections \ref{subsec:comparative_metrics_simple}-\ref{subsec:comparative_metrics_hybrid}, we take the models with the best metrics for a one-year forecast and run them for a long horizon. Table \ref{metricsLH} shows the comparative results of the one-year forecasts for all three cases and the extended forecasts for the Mines and UCD cases.

\newpage
\begin{longtable}[c]{ll | cc | lccc }
\caption{Comparative metrics of training and test sets of varying lengths for the selected models of  each case study. The GAM\textsubscript1 model has a spline cubic regression on the temperature regressor and the GAM\textsubscript2 model has a spline cubic regression on both the temperature and occupancy regressors. Test RMSE values are normalized by the mean of each data set, yielding the normalized root mean squared error (NRMSE). The {\it Peak} values are a percentage of the actual peak kW. The {\it Energy Use} values are a percentage of the actual annual energy use. } \label{metricsLH}\\
\multicolumn{2}{r}{} &
\multicolumn{2}{c}{Training set} &
\multicolumn{4}{c}{Test set} \\ \hline
        & Selected                  &              &             &               & NRMSE    & Peak     & Energy Use  \\ 
Case    & Model                     & Set length   & Adj-$R^2$   & Set length    & (\%)     & (\%)     & (\%)        \\ \hline

Mines   & GAM\textsubscript1+SARIMA & Nine years   & 0.8460      & One year     & 0.4349    & 104.1    & 99.56       \\
        & GAM\textsubscript1+SARIMA & One year     & 0.8332      & 11 years     & 9.091     & 93.22    & 84.49       \\ \hline

UCD     & MLR                       & Six years    & 0.6033      & One year     & 1.914     & 90.83    & 102.8       \\
        & GAM\textsubscript2+SARIMA & Six years    & 0.6371      & One year     & 0.4319    & 102.0    & 101.3       \\
        & MLR                       & One year     & 0.6320      & Six years    & 8.949     & 93.56    & 103.5       \\ 
        & GAM\textsubscript2+SARIMA & One year     & 0.7201      & Six years    & 9.634     & 93.87    & 103.8       \\ \hline  
        
Clemson & GAM\textsubscript2        & 1.2 years     & 0.8230      & One year     & 5.558     & 104.4    & 143.6       \\
        & GAM\textsubscript2+ARIMA  & 1.2 years     & 0.8335      & One year     & 6.765     & 92.20    & 93.23       \\ \hline

\end{longtable}

\begin{figure}[htb!]
	\centering
    \includegraphics[scale=0.46]{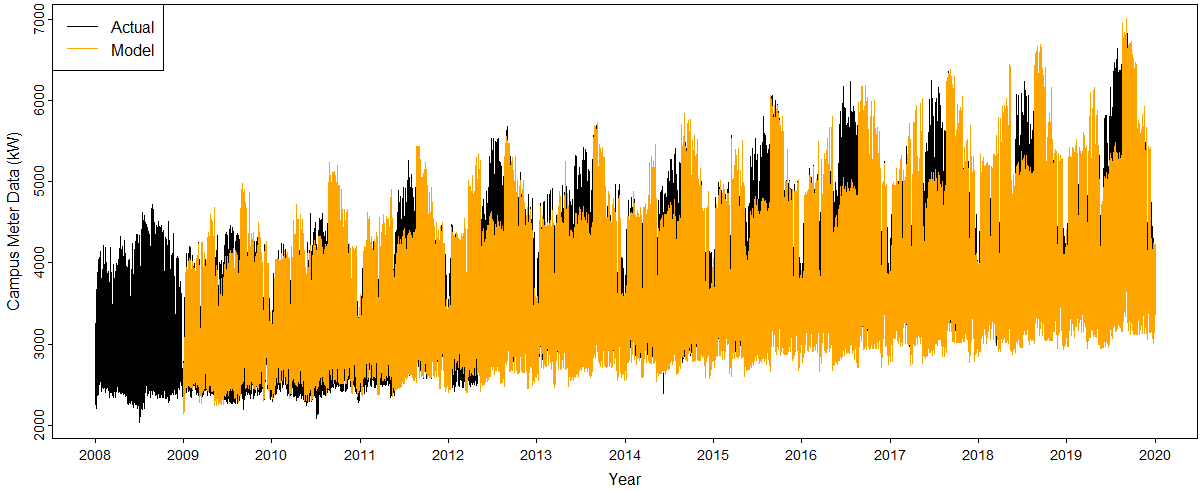}
    \caption{ Mines measured campus electrical demand data and GAM\textsubscript1+SARIMA forecast data for a one-year training and an 11-year test duration. } \label{fig:MinesLH} 
\end{figure}

For the Mines case, we take the selected model, GAM\textsubscript1+SARIMA, from the above statistical techniques using a nine-year training set and run the model with a shorter training period and correspondingly longer test length. The overall goal of this study is to forecast electric demand for longer than ten years. Because a single representative year of demand data is the common input for renewable energy optimization tools, we analyze a one-year training and an 11-year test set, which aligns with Mines' data availability.  Figure \ref{fig:MinesLH} shows the long horizon forecast, which yields a NRMSE value of 9.091\% of the actual mean and a peak value of 93.22\% of the actual.

While the mid-summer peak values fall below the actual data for the longer test periods, the peak values in August are relatively aligned with the actual data. The model captures the annual growth rates of monthly peak loads and the annual growth rates of overall energy consumption while preserving the daily, weekly, and month-to-month trends that occur within each year, or the seasonality, of the data. The model realistically represents the electric demand for a full range of weather and occupancy conditions.

\begin{figure}[htb!]
	\centering
    \includegraphics[scale=0.42]{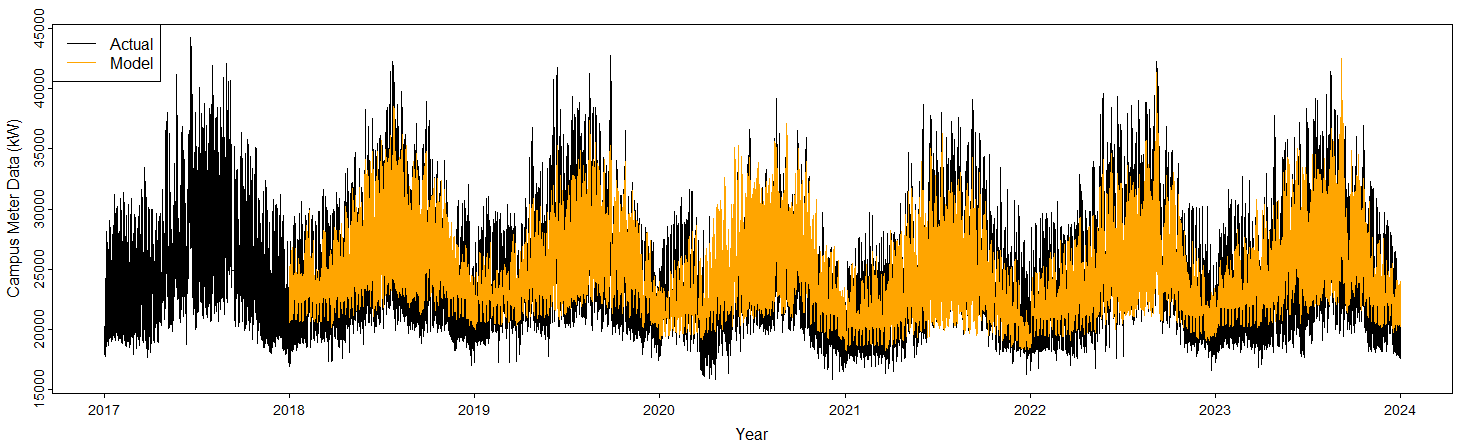}
    \caption{ Measured University of California Davis campus electrical demand data and MLR forecast data for a one-year training and a six-year test duration. } \label{fig:UCDMLR}
\end{figure}

For the UCD case, Table \ref{metricsLH} shows the metrics for the MLR and GAM\textsubscript2+SARIMA models. While the GAM\textsubscript2+SARIMA performs well in the six-year training, one-year test scenario, it does not translate well to a long horizon. Specifically,  the GAM\textsubscript2+SARIMA model has a test NRMSE of 0.4319\% for a short horizon, but this value increases to  9.634\% for the long horizon. The MLR model has a test NRMSE of 1.914\% for a short horizon,  which increases to 8.949\% for the long horizon; the latter value  is lower than that associated with the hybrid model. Figure \ref{fig:UCDMLR} shows the MLR six-year forecast. The forecast maximum is 91.13\% of the measured maximum. The total energy of the six-year test is 100.2\% of the total energy of six-year measured data. The residuals have minimal seasonality present. Therefore, we select the MLR as the most accurate predictive model for this data set.

\begin{figure}[htb!]
	\centering
    \includegraphics[scale=0.5]{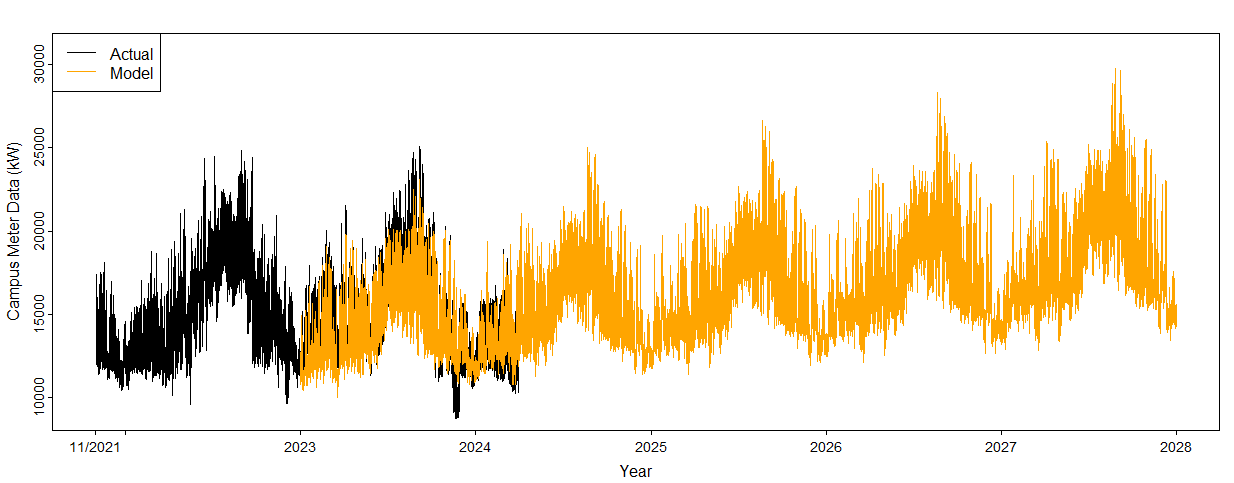}
    \caption{ Measured Clemson campus electrical demand data and GAM\textsubscript2+ARIMA forecast data for a 1.2-year training and a five-year forecast duration.} \label{ClemsonGAM2ARIMA}
\end{figure}

Table \ref{metricsLH} shows the metrics for both the GAM\textsubscript2 and GAM\textsubscript2+ARIMA models for Clemson. While the GAM\textsubscript2+ARIMA has a higher adjusted $R^2$ for the training set and test annual energy use closer to the measured value, the test NRMSE is lower and the peak value is closer for GAM\textsubscript2. Figure \ref{ClemsonGAM2ARIMA} shows the selected GAM\textsubscript2+ARIMA five-year forecast. Future exogenous variable data entails occupancy values that represent a continuation of past occupancy growth trends; Chowdhury et al. \citep{Chowdhury} provide future typical meteorological year (fTMY) weather data  from a multi-climate model. The fTMY used corresponds to a downscaled hourly surface temperature from the Intergovernmental Panel on Climate Change scenario SSP 2-4.5 from the 6th Coupled Model Intercomparison Project for  future years \citep{RIAHI2017}. The metrics for the selected GAM\textsubscript2+ARIMA model forecast for the first year are as follows: an NRMSE of 6.765\% of the actual mean, a maximum of 92.20\% of the measured maximum, and total energy of 92.23\% of the measured total energy use. With the test metrics quantifying the quality of only the first year of the forecast, the determination of the best model for subsequent years cannot be established. 

\newpage

\subsection{Machine learning model results}
\label{subsec:LSTM_forecast}

\indent The focus of this study is the framework for developing a forecast using classical statistical techniques, and, while artificial intelligence techniques have merit in this forecasting space, the interpretability of the forecast model is our priority. Nevertheless, we tune an LSTM model by an exhaustive search approach for hyper parameters and include it in the paper for critical comparison using the Mines and UCD case study data. Based on the short horizon exhibited in the Clemson data set and our corresponding inability to evaluate the quality of the forecast, combined with the need for the LSTM models to ``see" a relatively longer horizon (compared to the classical statistical models) in order to create a more accurate forecast, we do not subject this data set to the machine-learning models.

\subsubsection{Mines machine learning model results}
\label{subsec:LSTM_Mines}

\begin{figure}[htb]
	\centering
		\captionsetup{justification=centering}
	\includegraphics[scale=.4]{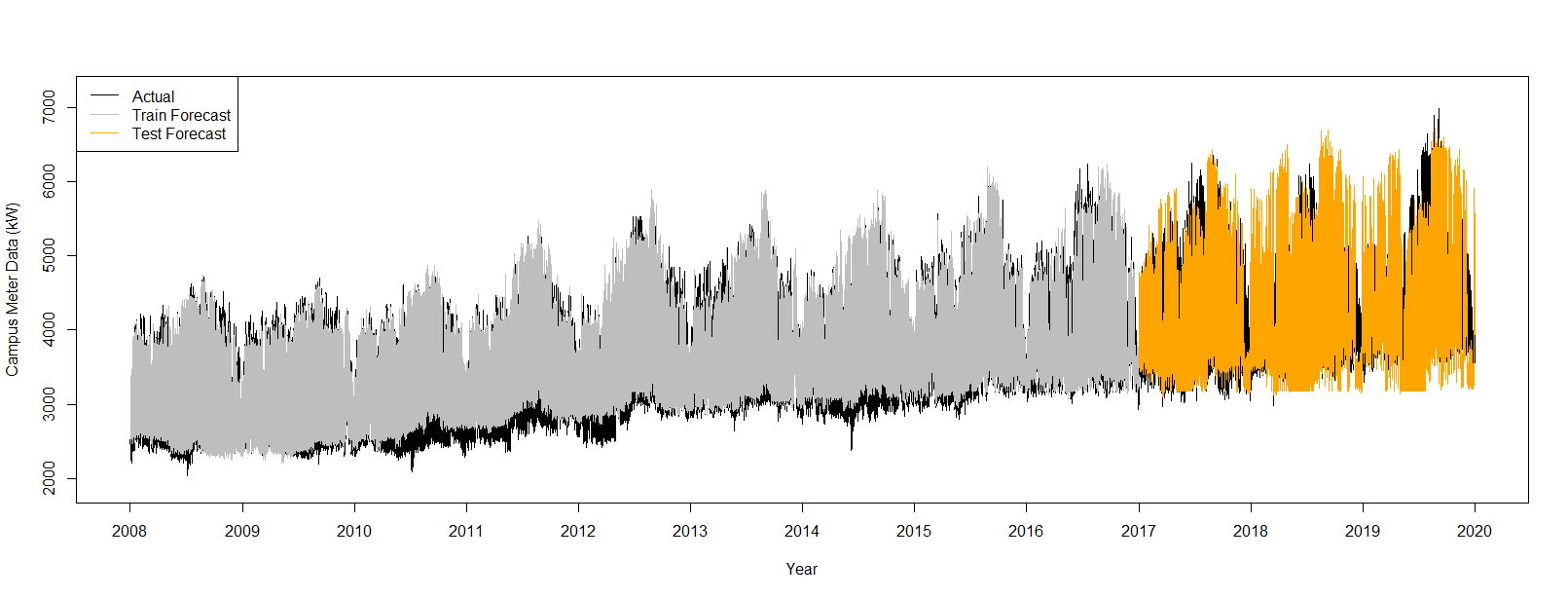} \\
	\caption{ Predicted values of the LSTM\textsubscript{2,b} model for a nine-year train and three-year test set for Mines.} \label{fig:LSTMMine}
\end{figure}

\indent The Mines LSTM models are trained and validated using the data on a nine-year train and half-year validation. The one-year predictions are made in a similar way. The models  have an input shape of 1 to 24 time steps and two to four features. Table \ref{metricsLSTM} shows the results of the four models that performed most favorably with respect to the three test metrics using the long training set, as well as the actual data and the best statistical model. The LSTM models for the nine-year training set with the highest adjusted $R^2$ are the LSTM\textsubscript{1,b} and LSTM\textsubscript{2,b} models with 0.88. Several of the adjusted $R^2$ results for the statistical models are greater than 0.88. The lowest RMSE for a one-year test for an LSTM model is 702.2 kW while the selected statistical model has an RMSE of 18.1 kW. Figure \ref{fig:LSTMMine} shows the selected LSTM\textsubscript{2,b} model for a long training set and short forecast.

\begin{longtable}[c]{lc | ccc | ccc}
\caption{Comparative metrics of training and one-year and three-year test sets of LSTM models on the Mines dataset. Model subscript 1: feature look-back of  $t-1$ and $t-24$. Model subscript 2: feature look-back of $t-1$, $t-2$,\ldots $t-24$. Model subscript a: features used are {\it Time-Step, Occupancy, Temperature}. Model subscript b: features used are {\it Time-Step, Occupancy, Temperature, Day Category}. Bold indicates the best results.} \label{metricsLSTM}\\
\multicolumn{2}{r}{Nine-year training set} &
\multicolumn{3}{c}{One-year test set} &
\multicolumn{3}{c}{Three-year test set} \\  \hline
Model                           & Adj-$R^2$      & RMSE          & Peak          & Energy Use    & RMSE          & Peak          & Energy Use  \\ 
                                &                & (kW)          & (kW)          & (MWh)         & (kW)          & (kW)          & (MWh)       \\ \hline
 Actual Data                    &  --            & --            & 6,353         & 36,481        & --            & 6,989         & 112,975     \\ \hline
 GAM\textsubscript1+SARIMA      & 0.85           & 18.10         & 6,612         & 36,322        & 15.55         & 7,385         & 112,567     \\ \hline
 LSTM\textsubscript{1,a}        & 0.83           & 736.3         & 6,000         & 37,464        & 787.7         & 6,587         & 112,360     \\
 LSTM\textsubscript{1,b}        & \textbf{0.88}  & 783.4         & 6,490         & 37,317        & 791.7         &\textbf{6,658} & 113,606     \\
 LSTM\textsubscript{2,a}        & 0.85           &\textbf{702.2} &\textbf{6,189} & 37,281        &\textbf{685.6} & 6,630         & 112,253     \\ 
 LSTM\textsubscript{2,b}        & \textbf{0.88}  & 800.1         & 7,358         &\textbf{37,035}& 834.7         & 7,407         & \textbf{112,510}\\ \hline
\end{longtable}

\indent Further, the proposed model selected for a long-horizon forecast consists of one LSTM layer with the output sequence of each layer as the input sequence of the next layer and consists of 200 neurons. We then use a dropout layer with a rate of 0.05. The LSTM layer and the dropout layer are each followed by a dense output layer, which possesses a Rectified Linear Unit (ReLU) as the activation function. The model is trained for 200 epochs with a batch size of 168.

\begin{figure}[htb]
	\centering
		\captionsetup{justification=centering}
	\includegraphics[scale=.4]{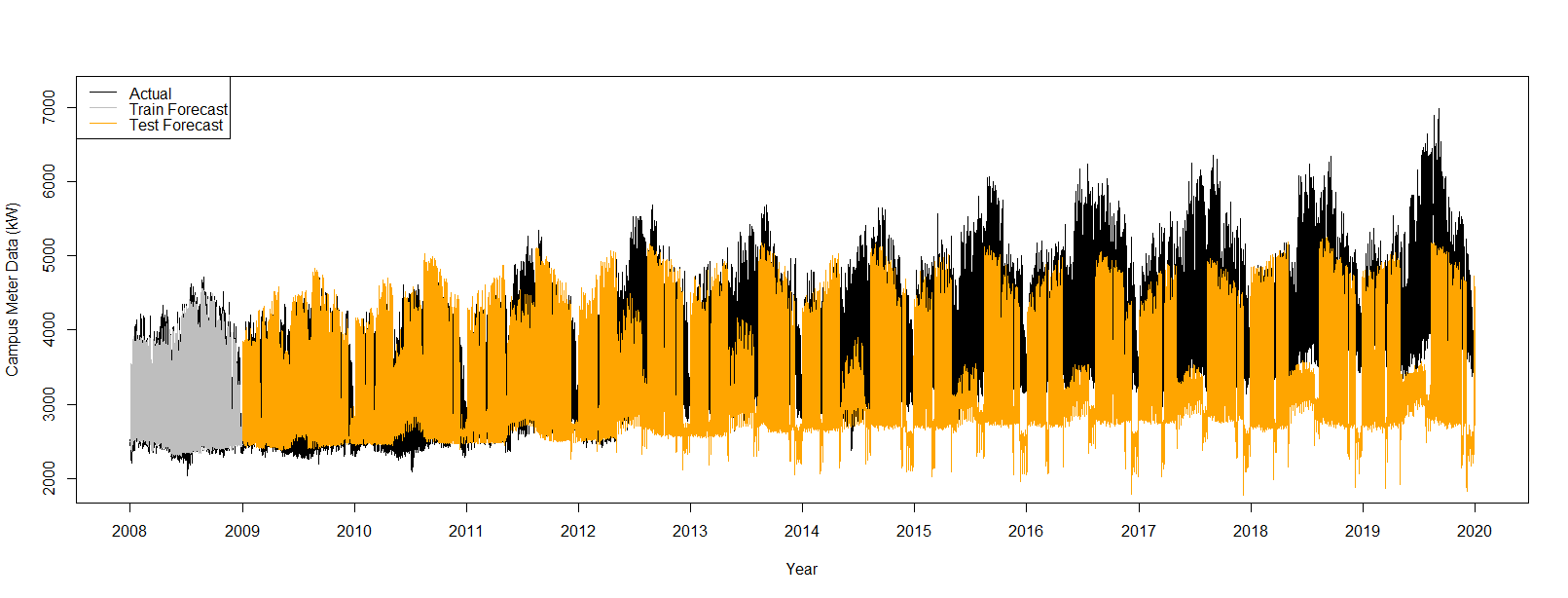} \\
	\caption{ LSTM predicted values of the LSTM\textsubscript{2,b} model for a one-year train and 11-year test set for Mines. } \label{fig:LSTMMines}
\end{figure}

\indent Table \ref{LSTMTrainTest} shows the comparative results of the extended forecasts for the LSTM\textsubscript{2,b} model. As the test periods get longer, the error of the forecast increases and is two to three times the error of the selected statistical model. The error compounds as the forecast values are developed from previous forecast values. Similar to the classical statistical model results shown in Table \ref{HybridTrainTestB}, the three- and five-year training sets result in higher test errors than the one-year training set. The long horizon LSTM\textsubscript{2,b} model, graphed in Figure \ref{fig:LSTMMines}, shows the increased error of the forecast year after year.

\newpage

\begin{longtable}[c]{ll | c | ccc}
\caption{Comparative metrics of Mines training and test sets of LSTM\textsubscript{2,b} and GAM\textsubscript1+SARIMA models for varying training and test data set lengths. Regressor look-back of $t-1$, $t-2$,... $t-24$ for {\it{Time-Step}, Occupancy, Temperature}, and {\it Day Category.}} \label{LSTMTrainTest}\\
\hline
 Model                     &                               & Adj-$R^2$            & RMSE (test)       & Peak           & Energy Use \\ 
                           &                               & (training)           & (kW)              & (kW)            & (MWh)  \\ \hline
 Actual Data (2017)        &                               &  --                  & --                & 6,353           & 36,481 \\ \hline
 LSTM\textsubscript{2,b}   & 9-year train, 1-year test     & 0.88                 & 765.8             & 6,387           & 37,315 \\
                           & 7-year train, 3-year test     & 0.87                 & 684.1             & 6,085           & 37,513 \\
                           & 5-year train, 5-year test     & 0.67                 & 1,430             & 7,742           & 42,652 \\
                           & 3-year train, 7-year test     & 0.83                 & 1,604             & 5,714           & 26,365 \\ \hline
 Actual Data (2019)        &                               &  --                  & --                & 6,989           & 39,059 \\ \hline
 LSTM\textsubscript{2,b}   & 1-year train, 11-year test    & 0.92                 & 1,364             & 5,257           & 30,625 \\ \hline
 GAM\textsubscript1+SARIMA & 1-year train, 11-year test    & 0.83                 & 346.0             & 6,515           & 33,002\\ \hline 
\end{longtable}

\newpage

\subsubsection{UCD machine learning model results}
\label{subsec:LSTM_UCD}

\indent The LSTM model for UCD is trained and validated using a one-year train and half-year validation with a six-year prediction. The model  has an input shape of 24 time steps and two features. Further, the proposed model selected for a long-horizon forecast consists of one LSTM layer with the output sequence of each layer as the input sequence of the next layer, and consists of 300 neurons. A dropout layer with a rate of 0.05 is used. The LSTM layer and the dropout layer are each followed by an output layer which is a dense layer with Rectified Linear Unit (ReLU) as the activation function. The model is trained for 400 epochs with a batch size of 216. 

\begin{figure}[htb!]
	\centering
    \includegraphics[scale=0.4]{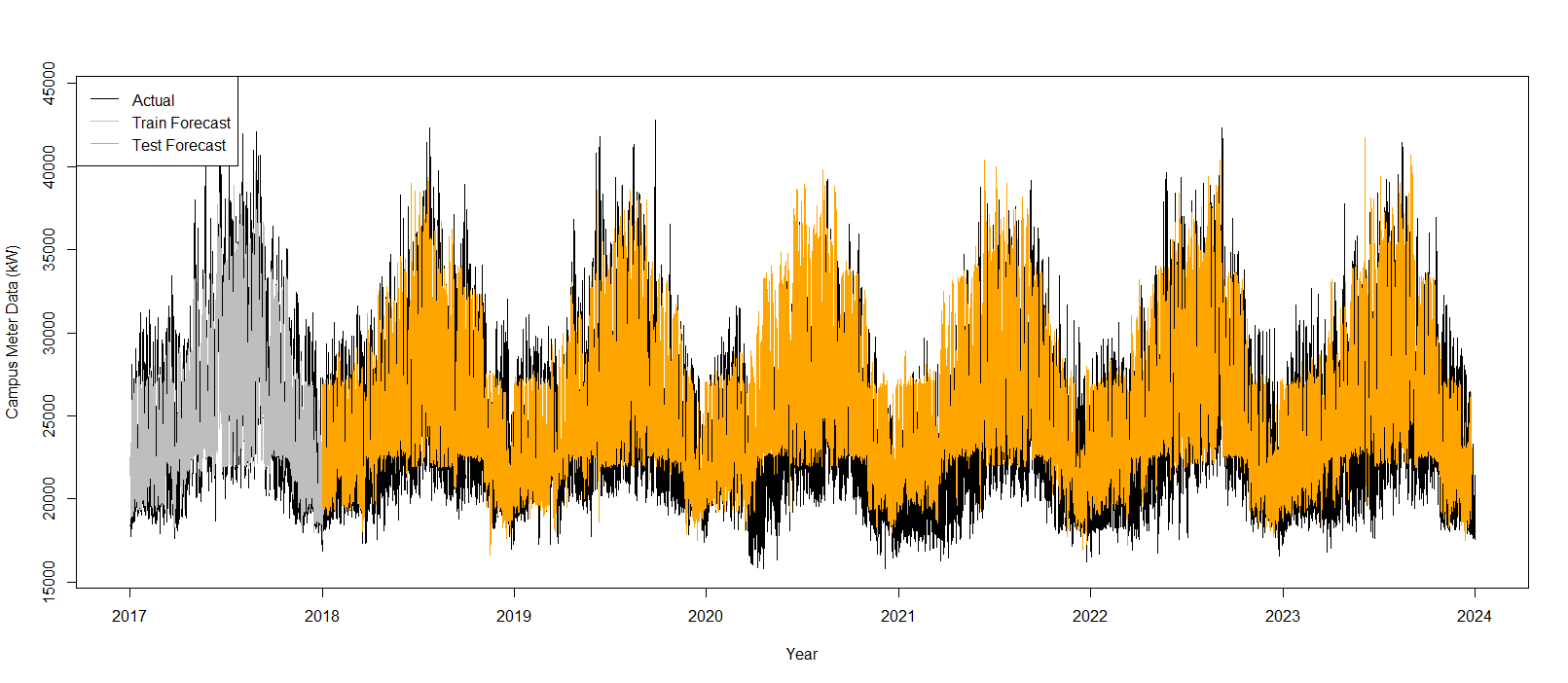}
    \caption{ Predicted values of the LSTM\textsubscript{2,a} model for a one-year train and six-year test set for UCD. } \label{fig:UCDLSTM}
\end{figure}

\indent Table \ref{LSTMUCD} shows the performance metrics of the selected LSTM\textsubscript{2,a} long-horizon forecast model for UCD compared to those associated with the selected statistical model. The six-year LSTM\textsubscript{2,a} forecast is shown in Figure \ref{fig:UCDLSTM}. The RMSE of the selected statistical model is 2,169 kW and is lower than the RMSE of the LSTM model. While the peak demand of the LSTM model is closer to that in the actual data, the annual energy use of the MLR model is closer to that observed in the actual data. A few traits of this data set make it a more likely candidate for an LSTM model than the Mines data set: the minimal change in the load profile from year to year, the consistency of the peak loads each year, and the high correlation between the energy demand and the temperature.

\begin{longtable}[c]{ll | c | ccc}
\caption{Comparative metrics of UCD training and test sets of LSTM\textsubscript{2,a} and MLR models. Regressor look-back of $t-1$, $t-2$,... $t-24$ for {\it{Time-Step}, Occupancy, Temperature,} and {\it Day Category.}} \label{LSTMUCD}\\
\hline
 Model                     &                               & Adj-$R^2$            & RMSE (test)       & Peak (2023)    & Energy Use (2023) \\ 
                           &                               & (training)           & (kW)              & (kW)            & (MWh)   \\ \hline
 Actual Data               &                               &  --                  & --                & 41,431          & 211,083 \\ \hline
 LSTM\textsubscript{2,a}   & 1-year train, six-year test   & 0.810                & 3,120             & 41,743          & 223,297 \\ \hline
 MLR                       & 1-year train, six-year test   & 0.632                & 2,169             & 39,597          & 219,818 \\ \hline 
\end{longtable}

%% file: conclusion.tex
\newpage

\section{Conclusions}
\label{conclusion}

\indent This study proposes a framework for statistical model selection to forecast long-horizon, hourly electric demand at the district energy level. We demonstrate the framework capabilities using the electric demand data collected from three campuses in the US: Colorado School of Mines, University of California Davis, and Clemson University. The Mines results reveal that the proposed GAM-SARIMA model is able to provide high-quality forecasts for electricity load demand over 11 years. The adjusted coefficient of determination (adjusted $R^2$) for modeling electric demand is 0.8332, implying that the proposed model could explain more than 83\% of the total variability within the training data. The NRMSE is 9.091\%, indicating that the average difference between the forecast and the mean measured data over an 11-year forecast is 9.091\%. The framework in this study produces a novel hybrid model type that incorporates classical linear regression, non-linear cubic spline coefficients, and seasonal time series modeling in order to represent a complex load profile for a district energy system. 

This framework for model development, when applied to two additional case studies, results in an MLR model and a GAM-ARIMA with regressors unique to the locations of UCD and Clemson, respectively. The resulting forecast model for UCD is an MLR model that yields a six-year forecast from a one-year training data set with a normalized root mean square error of 8.949\%. For Clemson, the forecast framework results in a GAM-ARIMA hybrid model with a one-year forecast from a 1.2-year training data set with a normalized root mean square error of 6.765\%. The advantage of the selected models is their preservation of accuracy for both long- and short-term forecasts, yielding a forecast from one year of training data. We additionally develop machine learning models, but their performance across our metrics of choice was not competitive with that of the classical statistical models; additionally, the latter affords us the advantage of interpretability of the regressors invoked.

\indent An extension to this study develops a long-horizon, hourly fidelity renewable energy optimization model in which we can incorporate our electric demand forecasts. At the time of this writing, the available techno-economic optimization tools are unable to accommodate long horizons. Future work may also include forecasting under uncertainty and with data sets from different settings, e.g., that might see a growth in load as a result of  the electrification of natural-gas equipment and a significant amount of electric vehicle  charging.

%% file: acknowledgements.tex
\section*{Acknowledgements}
\label{acknowledgements}

\indent The authors would like to thank the Colorado School of Mines’ Facilities Office (especially Sam Crispin, Michael Willey, and Mike Bowker) for providing campus data and technical information in support of this project. The authors would like to thank Joseph Yonkoski at the University of California Davis and Bret McCarley and Thomas Stuttles of Clemson University for compiling and sharing electric demand data from their universities. The authors also acknowledge assistance from Dr. Alexander Zolan from the National Renewable Energy Laboratory, and Kathleen Tomon, Dr. Madeline Macmillan, Dr. Samy Wu Fung, and Dr. Daniel McKenzie from the Colorado School of Mines for the insights provided on long-horizon forecasting. Finally, we acknowledge two anonymous reviewers whose comments helped to strengthen the paper.

%% file: appendix.tex
\newpage
\newpage
\newpage
\section{Mathematical Formulation}
\label{appendix}

\indent We present the mathematical formulation for the selected hybrid forecast model GAM\textsubscript1+SARIMA for the Colorado School of Mines campus in Table \ref{tab:model_formulation}. The exogenous variable temperature has a spline cubic regression formula for Mines. The GAM\textsubscript2+ARIMA model for Clemson differs from the Mines formulation in two ways; the exogenous variable occupancy also has a spline cubic regression formula and the non-seasonal and seasonal operators have a different orders. Equation \eqref{eq:gam1} shows the GAM\textsubscript1 model formulation.

\begin{longtable}[c]{llll}
\caption{Notation of the coefficients and variables for the final model formulation used in equations \eqref{eq:gam1}-\eqref{eq:residualsarima}.} \label{tab:model_formulation} \\
 Coefficient          &                                                  &                 & Units \\ \hline
 $\alpha$             & intercept for GAM\textsubscript1 model           &                      & [kW]  \\
 $\beta_1$            & coefficient of the exogenous variable $\it x_1$  &  & [-] \\
 $\beta_2$            & coefficient of the exogenous variable $\it x_2$  &  & [-] \\
 $\it{f}_3$           & smoothed function of the exogenous variable $\it x_3$  &   & [-] \\
 $\beta_4$            & coefficient of the exogenous variable $\it x_4$  &  & [-] \\
 $\beta_5$            & coefficient of the exogenous variable $\it x_5$  &  & [-] \\
 $\beta_6$            & coefficient of the exogenous variable $\it x_6$  &  & [-] \\
 $\beta_7$            & coefficient of the exogenous variable $\it x_7$  &  & [-] \\
 $\phi_1$             & first order non-seasonal AR operator             &  & [-] \\
 $\phi_2$             & second order non-seasonal AR operator            &  & [-] \\
 $\phi_3$             & third order non-seasonal AR operator             &  & [-] \\
 $\phi_4$             & fourth order non-seasonal AR operator            &  & [-] \\
 $\phi_5$             & fifth order non-seasonal AR operator             &  & [-] \\
 $\Phi_1$             & first order seasonal AR operator                 &  & [-] \\
 $\theta_1$           & first order non-seasonal MA operator             &  & [-] \\
 & & \\
 Variables            &                                                 &               & Units \\ \hline
 $\it r_t$            & residuals from GAM\textsubscript1 model at time $\it t$         &  & [kW]      \\
 & & & \\
 \multicolumn{2}{l}{Exogenous Variables} & \multicolumn{1}{r}{} & \multicolumn{1}{l}{Units} \\    \hline
 $\it x_{1t}$         & time step \textit{t}                           &                & [hour] \\
 $\it x_{2t}$         & occupancy at time \textit{t}                   &                & [number of people] \\
 $\it x_{3t}$         & temperature at time \textit{t}                 &                & [$^\circ C$] \\
 $\it x_{4t}$         & day category at time \textit{t}                &                & [-] \\
 $\it x_{5t}$         & class binary at time \textit{t}                &                & [-] \\
 $\it x_{6t}$         & cosine term at time \textit{t}                 &                & [-] \\
 $\it x_{7t}$         & sine term at time \textit{t}                   &                & [-] \\
\end{longtable}

\begin{equation} \label{eq:gam1}
    {y}_t = {\alpha} + {\beta}_1x_{1t} + {\beta}_2 x_{2t}  + f_3x_{3t} + {\beta}_4 x_{4t}  + {\beta}_5 x_{5t}  + {\beta}_6 x_{6t}  + {\beta}_7 x_{7t}  + {\epsilon} \;\; \forall t\in\mathcal{T}
\end{equation}

\indent From there, the residuals from the GAM\textsubscript1 model are calculated using equation \eqref{eq:gamresiduals}. 

\begin{equation} \label{eq:gamresiduals}
    r_t = y_t - \hat{y}_t  \;\; \forall t\in\mathcal{T}
\end{equation}

The SARIMA orders for $(p,d,q)(P,D,Q)m$, as described in Section \ref{subsec:time_series}, are $(5,1,1)(1,0,0)24$. In the SARIMA formulation, $B$ is the backshift operator, with the seasonal duration of 24 time steps shown in equation \eqref{eq:backshift}. The formulation for the SARIMA model is shown in equation \eqref{eq:residualsarima} with $\varepsilon_t$ as the white noise value at period $\it{t}$. 

\begin{equation} \label{eq:backshift}
        B^{24}y_t=y_{t - 24}
\end{equation}
\begin{equation} \label{eq:residualsarima}
   (1-\phi_1B^1-\phi_2B^2-\phi_3B^3-\phi_4B^4-\phi_5B^5)(1-\Phi_1B^{24})(1-B)r_t = (1+\theta_1B^1)\varepsilon_t  \;\; \forall t\in\mathcal{T}
\end{equation}

%% file: appendixB.tex
\section{Detailed Results from All Case Studies}
\label{appendixB}

\indent This Appendix details the results from the application of the statistical model selection framework and is intended to supplement the results in Section \ref{sec:results_test_case}. The supplemental Mines results describe the initial data analysis, exogenous variable details, and the comparison of various long-horizon forecasts with different training set lengths. The UCD supplement presents all three categories of classical statistical models using actual, rather that normalized, values. For Clemson, the supplemental information focuses on comparing multiple forecasts, rather than providing only the selected model in Section \ref{subsec:long_horizon_forecast}.

\subsection{Colorado School of Mines}
\label{subsec:AppendixBMines}

\begin{figure}[htb!]
	\centering
    \includegraphics[scale=0.55]{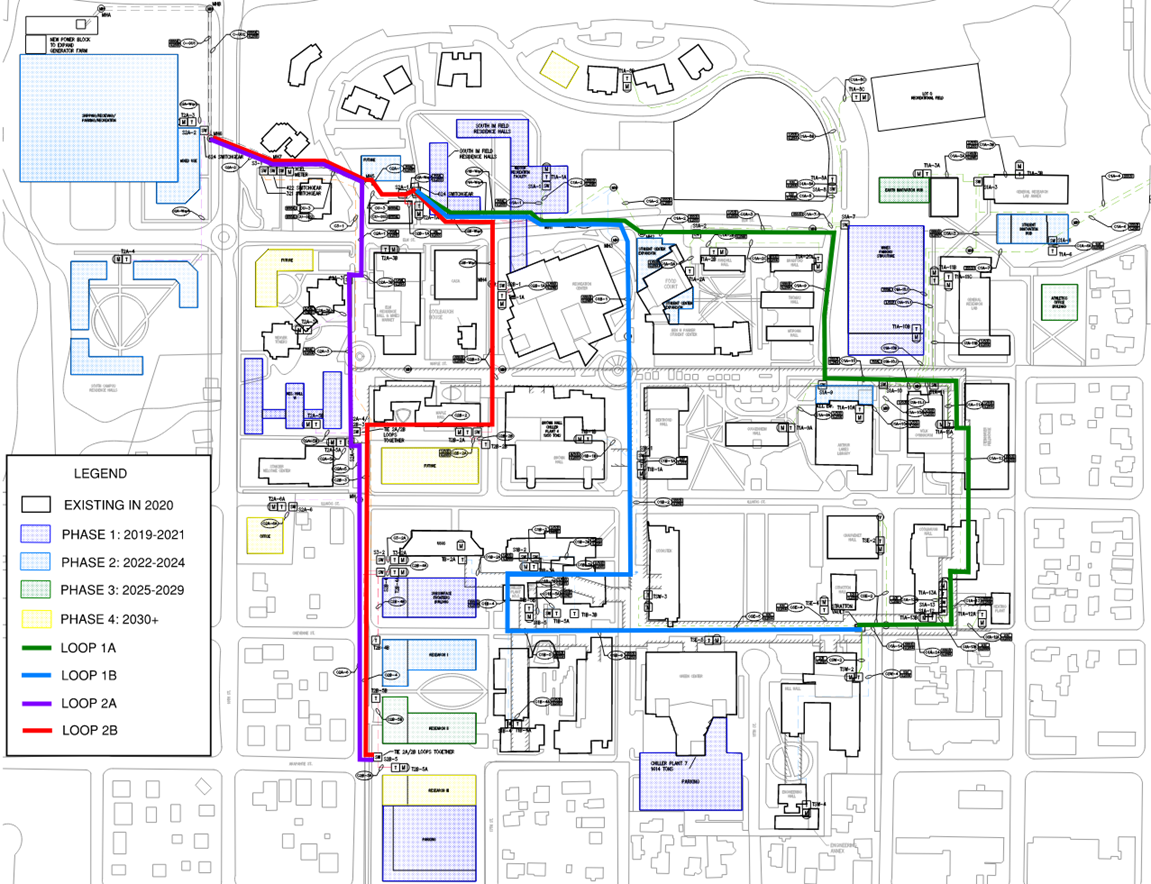}
    \caption{Existing electrical site plan with future buildings. } \label{fig:CampusMapB}
\end{figure}

\indent We aggregate the primary electric meter demand data from the Mines campus, shown in Figure \ref{fig:CampusMapB}, from its original fidelity of 15-minute intervals, into hours by taking the sum of the four 15-minute intervals for each hour. We check the hourly campus meter data for zeros and NAs before log-transforming it. We replace zero values, which would correspond either to a power or metering outage, with the mean of the hour before and the hour after the value. In the event of an outage lasting over one hour, we replace the zero values with the scaled hourly data from the previous day; the demand values before and after the outage compared to the reference day determine the scaling factor. Figure \ref{fig:CampusLoad} shows the hourly aggregated Mines meter data at the primary Xcel Energy utility meter.
 
\begin{figure}[htb!]
	\centering
	\subfloat[Campus Meter Data]{\includegraphics[scale= 0.58]{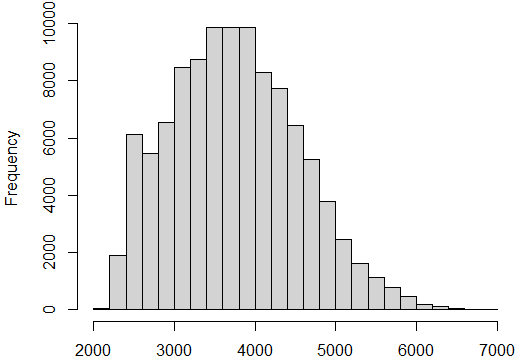} \label{sub:CHistoB}}
    \subfloat[Log-transformed Campus Meter Data]{\includegraphics[scale= 0.58]{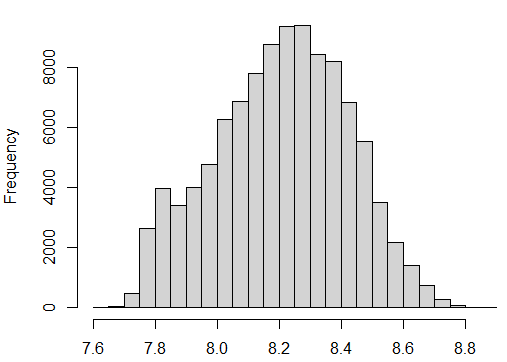} \label{sub:LHistoB}} \\
	\caption{Comparison of histograms before and after log transformation}
	\label{fig:HistosB}
\end{figure}

\indent We check the meter data for normality using the Kolmogorov-Smirnov test; the p-value of the test is much smaller than 0.05, which indicates that the data is non-normally distributed. Figure \ref{sub:CHistoB} shows that the marginal distribution of the data has a large range. To reduce the fluctuation of the data values and the skewness and to exaggerate the periodic structure, we perform a logarithmic transformation. Figure \ref{sub:LHistoB} shows the stabilized variance of the log-transformed time-series data set.

\begin{figure}[htb!]
	\centering
    \includegraphics[scale=0.46]{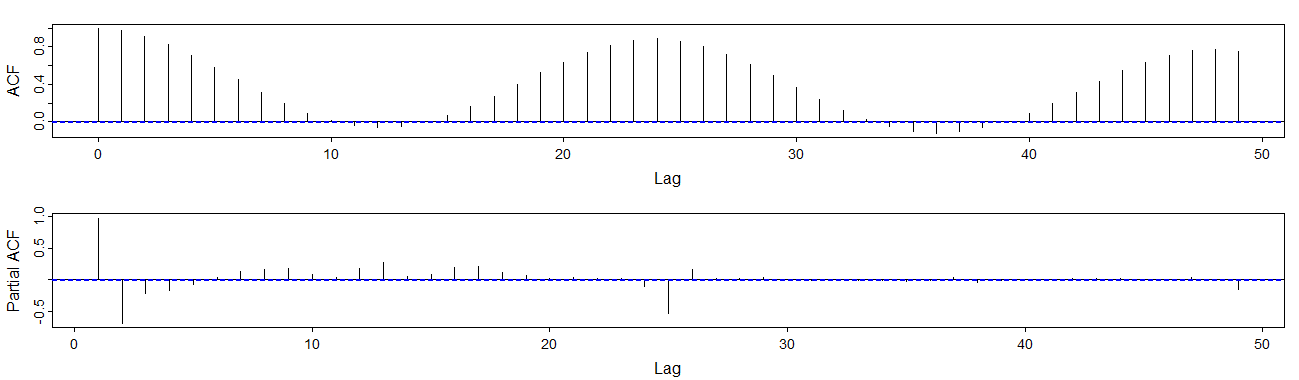}
    \caption{Plots of the autocorrelation function (ACF) and partial autocorrelation function (PACF) of the log-transformed campus meter data indicate a 24-hour lag.} \label{fig:CMACFB}
\end{figure}

\indent Figure \ref{fig:CMACFB} shows the seasonal nature of the actual meter data in an autocorrelation function plot and partial autocorrelation function plot. The $y$-axis represents the correlation value between a data point $y$\textsubscript{t} and data point $y$\textsubscript{t+lag}; a value of zero indicates no correlation. The autocorrelation function plot suggests that \begin{math}y\textsubscript{t}\end{math} has a correlation of around 0.8 with \begin{math}y\textsubscript{t+24}\end{math}, \begin{math}y\textsubscript{t+48}\end{math}, and so on.  The magnitude of the correlation is maintained. The seasonality captured here informs the  regressors and time series orders.

\textbf{Selected exogenous variables}

\begin{figure}[htb!]
	\centering
		\captionsetup{justification=centering}
	\includegraphics[scale=.50]{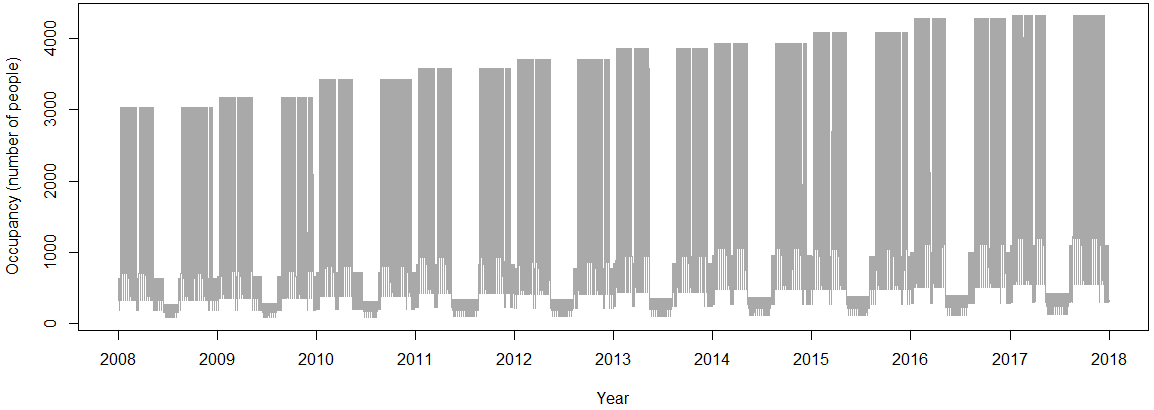}
\\
	\caption{ Hourly campus occupancy data for Mines }\label{fig:FTEB}
\end{figure}

\begin{figure}[htb!]
	\centering
		\captionsetup{justification=centering}
	\includegraphics[scale=.50]{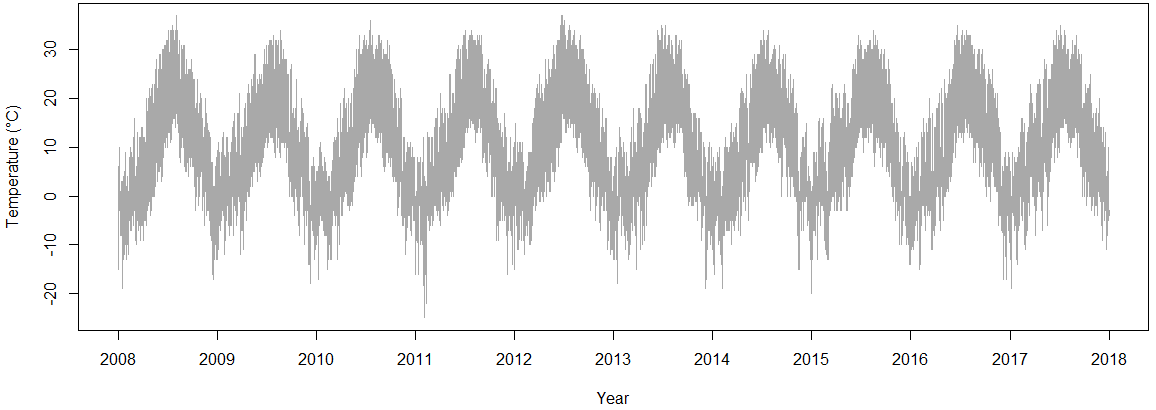} \\
	\caption{ Hourly actual meteorological year outdoor air temperature for Golden, Colorado} \label{fig:tempB}
\end{figure}

\indent Figure \ref{fig:FTEB} shows the estimated hourly occupancy of all of the buildings on the Mines campus. This encompasses the historical full-time-equivalent student, faculty, and staff information available for each school year, as well as the number of students living on campus. Figure \ref{fig:tempB} shows the hourly actual meteorological year outdoor air temperature for Golden, Colorado. Note that the occupancy is increasing year-after-year, while the outdoor air temperature peak values are not; these characteristics influence the correlation to the electric demand.
\vspace{-0.2cm}

\begin{figure}[htb!]
	\centering
		\captionsetup{justification=centering}
	\includegraphics[scale=.51]{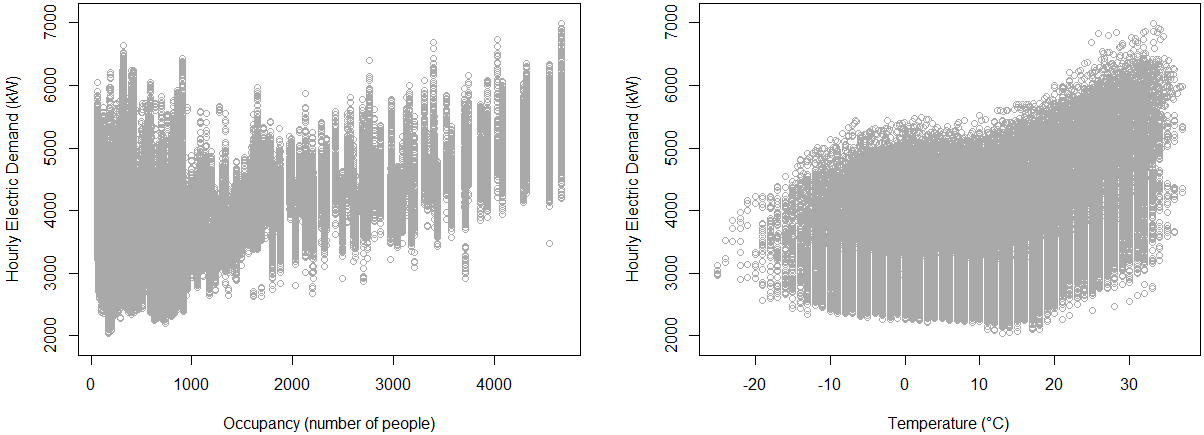} \\
	\caption{Scatterplots of Occupancy versus Meter Data and Temperature vs Meter Data to show linearity. } \label{fig:scatterB}
\end{figure}

\indent Figure \ref{fig:scatterB} provides scatterplots of the electric load demand versus the campus occupancy, and electric load demand versus the outdoor air temperature. The estimated correlations between electricity load demands with occupancy and outdoor air temperature are 0.470 and 0.436, respectively. While the correlations are close in value, the scatterplots show a linear correlation with occupancy and a non-linear correlation with outside temperature. The increase in electric demand with higher temperatures is due to Mines using electricity for cooling, but natural gas for heating. To develop the most accurate model for the electricity load demand, we must add more  significant exogenous factors than occupancy and temperature. For example, the occupancy rates are estimated, but whether class is in session or it is a weekend day are determined by the academic calendar. Adding categorical  variables provides an explanation for days with similar outdoor air temperatures that might have different electric demands. 

\indent We define the additional initial exogenous variables for Mines as follows: {\it Humidity} is the actual meteorological year outdoor relative humidity hourly data set for Golden, Colorado. {\it Total Building Area} is the  total building area of the campus as buildings are connected to the primary electric meter over time. {\it Energy Use Intensity} is the  amount of energy used per square foot annually. {\it Day Category} is defined as the following six categories: spring and fall semester weekday, spring and fall semester weekend, summer weekday, summer weekend, holiday break weekday, and holiday break weekend. These day categories capture the behavior of thermostat setbacks, on-campus residents, and non-class-related events. {\it Class Binary} assumes a value of one when it is a standard class day, and zero for all other days. There are many days during the year on which most faculty, staff, and administration are on campus but classes are not held.  

\indent This study uses LASSO regression to select the appropriate exogenous variables. LASSO is a modification of linear regression that minimizes the complexity of the model by limiting the sum of absolute values of the model coefficients \citep{Deng}. We perform k-fold cross-validation to find the optimal $\lambda$ value (see Equation \ref{eq:LASSO}) of 0.000492 to minimize the mean-square error. Then, using the best $\lambda$, we determine the LASSO regression coefficients. Outdoor relative humidity, total building area, and energy use intensity return a coefficient of zero in the LASSO regression; hence, we omit them from the model.

\bigskip
\textbf{Comparative metrics of exogenous variable regression models}

\indent An MLR is applied to the remaining exogenous variables after the LASSO selection of {\it{Time-Step}, Occupancy, Temperature, Day Category, Class Binary, Cosine Term,} and {\it Sine Term}. To further improve the fit of the linear regression, we include non-linear coefficients for exogenous variables with non-linear correlation in a GAM model. We train a matrix of combinations of spline and spline cubic regression functions on the exogenous variables. The best GAM model fit configurations are with GAM\textsubscript1, which has a spline cubic regression on the temperature regressor, and GAM\textsubscript2, which has a spline cubic regression on both the temperature and occupancy regressors.

\begin{figure}[htb]
	\centering
		\captionsetup{justification=centering}
	\includegraphics[scale=.49]{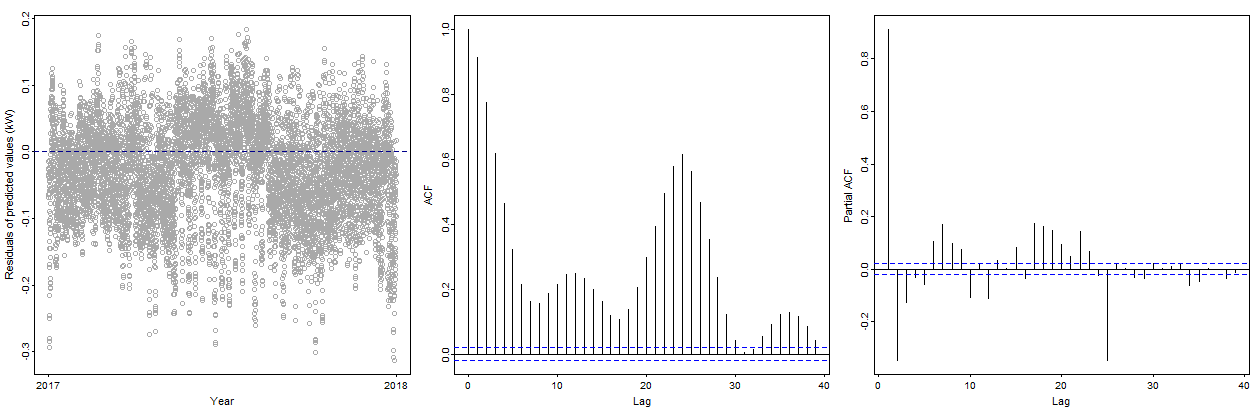} \\
	\caption{ Residuals of the predicted values of the log GAM\textsubscript1 model and the log campus meter data. The autocorrelation function and partial autocorrelation function show the seasonality remaining in the residuals of the log GAM\textsubscript1 model. } \label{fig:GAMresidualB}
\end{figure}

\indent Table \ref{metrics1B} compares the accuracy of each analyzed simple technique to model and forecast electric demand for the nine-year training and test data sets. We use both a one-year and a three-year test data set; while a single year is standard, the three-year test better reflects some of the higher peaks as a result of a changed metering procedure on the campus. For the training data set, GAM\textsubscript1 has the highest adjusted $R^2$ value. And, while GAM\textsubscript1 has the best RMSE values on the one-year test data set, GAM\textsubscript2 has the best RMSE on the three-year test data set.  The residuals, autocorrelation function plot, and partial autocorrelation function plot for the GAM\textsubscript1 model, shown in Figure \ref{fig:GAMresidualB}, indicate 24-hour trends that are not captured with the model. The results are similar for the residuals of all three simple models. The GAM\textsubscript1 has the lowest NRMSE and closest annual energy use and GAM\textsubscript2 has the closest peak value, so we examine both models  further in the following section.

\newpage
\begin{longtable}[c]{lcc | ccc | ccc}
\caption{Comparative metrics of training, and one-year and three-year test, sets of simple exogenous variable  models. The GAM\textsubscript1 model has a spline cubic regression on the temperature regressor and the GAM\textsubscript2 model has a spline cubic regression on both the temperature and occupancy regressors. Bold indicates the best results.} \label{metrics1B}\\
\multicolumn{3}{r}{Nine-year training set} &
\multicolumn{3}{c}{One-year test set} &
\multicolumn{3}{c}{Three-year test set} \\ \hline
Model      && Adj-$R^2$     & RMSE      & Peak       & Energy Use & RMSE      & Peak   & Energy Use  \\ 
           &&               & (kW)      & (kW)       & (MWh)      &  (kW)     & (kW)   & (MWh)       \\ \hline
Actual Data&&  --           & --        & 6,353      & 36,481     & --        & 6,989  & 112,975 \\ \hline
MLR        && 0.7908        & 133.6     & 6,123      & 37,651     & 158.5     & \textbf{6,768}  & 117,141 \\
GAM\textsubscript1       && \textbf{0.8830} & \textbf{122.7} & 6,845  & \textbf{37,520}  & 171.9     & 7,791  & 117,494 \\
GAM\textsubscript2       && 0.8682        & 141.3     & \textbf{6,570}      & 37,719 & \textbf{56.97}& 7,447  & \textbf{114,472} \\ \hline
\end{longtable}

\bigskip
\textbf{Comparative metrics of time series models}

\indent Linear and non-linear regression models that use only exogenous variables to forecast electric demand do not adequately capture the seasonal behavior of the actual campus meter data; therefore, we also evaluate time series models. Our analysis includes ARIMA and SARIMA, but alone, they do not capture the correlation with the exogenous variables, and, thus, we do not portray the results here. Table \ref{metrics2B} compares the accuracy of a sample of analyzed non-seasonal and seasonal time-series models and forecast electric demand for the nine-year training, and one-year and three-year test data sets. These models have flat behavior, failing to capture the peak values of the actual meter data. With these shortcomings of each classical statistical forecast model used independently, we evaluate the hybrid models in the following section.

\begin{longtable}[c]{lcc | ccc | ccc}
\caption{Comparative metrics of training and one-year and three-year test sets of time series models.}\label{metrics2B}\\
\multicolumn{3}{r}{Nine-year training set} &
\multicolumn{3}{c}{One-year test set} &
\multicolumn{3}{c}{Three-year test set}   \\ \hline
Model                   && Adj-$R^2$     & RMSE      & Peak       & Energy Use & RMSE      & Peak   & Energy Use  \\ 
                        &&               & (kW)      & (kW)       & (MWh)      &  (kW)     & (kW)   & (MWh)       \\ \hline
Actual Data             &&  --           & --        & 6,353      & 36,481     & --        & 6,989  & 112,975 \\ \hline
ARIMA(0,1,5)            && 0.9669        & 1,258     & 3,109      & 27,242     & 2,179     & 3,110  & 81,727 \\
ARIMA(2,1,1)            && 0.9660        & 1,322     & 3,076      & 26,580     & 2,290     & 3,076  & 79,739 \\
SARIMA(0,1,5)(0,0,1)24  && 0.9781        & 1,250     & 3,202      & 27,326     & 2,165     & 3,202  & 81,979 \\
SARIMA(2,1,1)(0,0,1)24  && 0.9778        & 1,306     & 3,159      & 26,750     & 2,261     & 3,159  & 80,249 \\ \hline
\end{longtable}

\newpage
\textbf{Comparative metrics of hybrid models}

\indent Hybrid models can capture daily trends better than simple regression models. Table \ref{metrics3B} compares the accuracy of each analyzed hybrid technique to model and forecast electric demand for the nine-year train and one-year and three-year test data sets. For the training data set, MLR+ARIMA has the highest adjusted $R^2$ value. For the one-year and three-year test data sets, GAM\textsubscript1+SARIMA has lower RMSE values than MLR+ARIMA and all other models. The GAM\textsubscript1+SARIMA model captures and exceeds the peak values and closely aligns with the total energy use. As noted in Section \ref{subsec:evaluating_model_performance} as a metric of evaluation, the time-of-day of the peak values for all models is 1:00pm, which aligns with the actual meter data. Overall, GAM\textsubscript1+SARIMA provides the most accurate results among all analyzed techniques in longer test data sets. The formulation for the GAM\textsubscript1+SARIMA model is provided in \ref{appendix}.

\begin{longtable}[c]{lc | ccc | ccc}
\caption{Comparative metrics of training and one-year and three-year test sets of hybrid statistical models. Bold indicates the best results.} \label{metrics3B}\\
\multicolumn{2}{r}{Nine-year training set} &
\multicolumn{3}{c}{One-year test set} &
\multicolumn{3}{c}{Three-year test set} \\ \hline
Model                                      & Adj-$R^2$      & RMSE          & Peak          & Energy Use    & RMSE          & Peak          & Energy Use  \\ 
                                           &                & (kW)          & (kW)          & (MWh)         & (kW)          & (kW)          & (MWh)       \\ \hline
 Actual Data                               &  --            & --            & 6,353         & 36,481        & --            & 6,989         & 112,975 \\ \hline
 MLR+ARIMA(0,0,1)                          & \textbf{0.9746}& 132.5         & 7,967         & 45,221        & 156.6         & 6,733         & 117,089 \\
 MLR+SARIMA(0,0,1)(1,0,0)24                & 0.9495         & 140.3         & \textbf{6,222}& 37,710        & 149.7         & \textbf{6,861} & 116,909 \\
 GAM\textsubscript1+ARIMA(5,1,1)           & 0.8508         & 454.0         & 5,911         & 32,708        & 405.3         & 6,786         & 102,325 \\
 GAM\textsubscript2+ARIMA(0,1,5)           & 0.8917         & 500.1         & 5,578         & 32,327        & 565.7         & 6,382         & 98,110  \\
 GAM\textsubscript1+SARIMA(5,1,1)(1,0,0)24 & 0.8460         & \textbf{18.10}& 6,612         & \textbf{36,322}& \textbf{15.55}& 7,385        & \textbf{112,567} \\
 GAM\textsubscript2+SARIMA(0,1,5)(1,0,0)24 & 0.8720         & 733.0         & 5,221         & 30,060        & 877.7         & 5,752         & 89,909  \\ \hline
\end{longtable}

\indent Figure B.\ref{fig:GAMSARIMA}b shows the forecast results for the GAM\textsubscript1+SARIMA model for a one-year test dataset, specifically the model alignment with actual data in representative monthly, weekly, and daily load demand patterns. The month of August historically has the highest electric demand values and represents the essential load for optimal sizing of renewable energy technologies.  In addition, for most district energy systems, including at Mines, the utility billing demand charges are based on the highest demand in any given month on a non-holiday weekday between 2:00pm and 6:00pm. If the highest forecast demand value in each month is lower than the actual value, models that evaluate the cost savings of a set of renewable energy technologies will provide a conservative estimate.  

\indent The one-year and three-year RMSE for the GAM\textsubscript1+SARIMA model shows an improvement in the forecast accuracy and in the total energy use estimate over the GAM\textsubscript1 model given in Table \ref{metrics1B}.

\bigskip
\textbf{Long-horizon electric demand forecast}

\begin{figure}[htp!]
	\centering
        \captionsetup{justification=centering}
	\subfloat[Measured campus electrical demand data and GAM\textsubscript1+SARIMA forecast data for a seven-year training and a three-year test duration.]{\includegraphics[scale=.38]{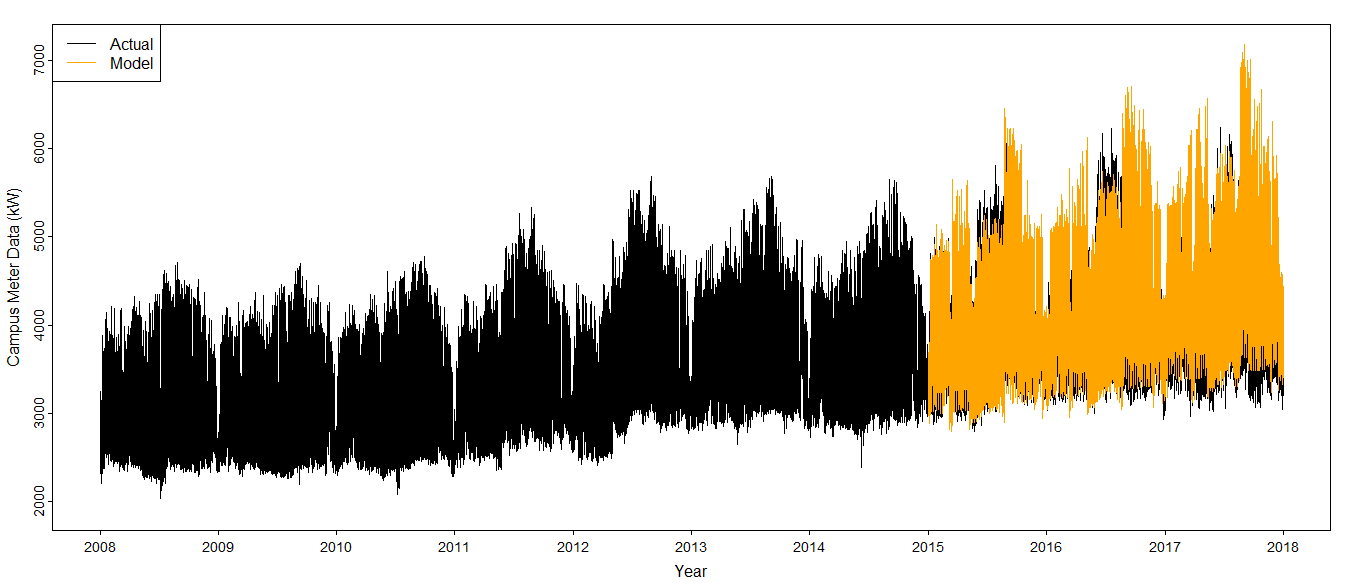} \label{sub:7-3yearB}} \\
    \subfloat[Measured campus electrical demand data and GAM\textsubscript1+SARIMA forecast data for a one-year training and a nine-year test duration.]{\includegraphics[scale=.38]{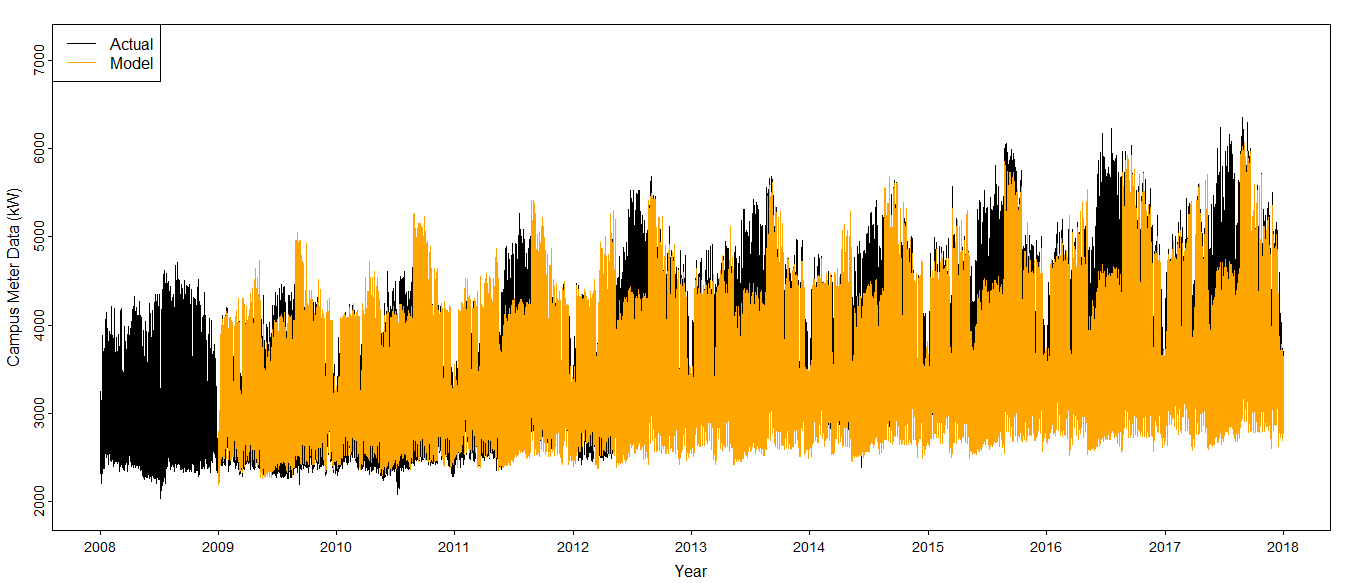} \label{sub:10yearB}}\\   
	\subfloat[Measured campus electrical demand data and GAM\textsubscript1+SARIMA forecast data for a one-year training with an 11-year test duration.]{\includegraphics[scale= 0.42]{Mines_12YEAR.png} \label{sub:12yearB}}\\
	\caption{Measured campus electrical demand data and GAM\textsubscript1+SARIMA forecast data for selected training and test duration.} 
	\label{fig:ExtendedB}
\end{figure}

\indent The overall goal of this study is to forecast electric demand for longer than ten years. We take the selected model, GAM\textsubscript1+SARIMA, from the above statistical techniques using a nine-year training set and run the model with shorter training periods and  correspondingly longer test lengths. Because a single representative year of demand data is the common input for renewable energy optimization tools, we  change to a one-year training and an 11-year test, which aligns with the available data from Mines. Table \ref{HybridTrainTestB} shows the comparative results of the extended forecasts. Figure \ref{sub:7-3yearB} and \ref{sub:10yearB} demonstrate that the reduction of the test set length to seven years and one year, respectively, yield RMSE values of 87.54 kW and 284.3 kW, respectively. Figure \ref{sub:12yearB} depicts an extension of the forecast horizon to 11 years, with a corresponding RMSE of 346.0 kW. 

\newpage
\begin{longtable}[c]{lc | ccc}
\caption{Comparative metrics of training and test sets of GAM\textsubscript1+SARIMA models for a varying training and test data set lengths. Bold indicates the results of the models shown in Figures \ref{sub:7-3yearB}-\ref{sub:12yearB}. \label{HybridTrainTestB}}\\
\hline
 Model                         & Adj-$R^2$  & RMSE (test)     & Peak             & RMSE (test over 5,000 kW) \\ 
                               & (training) & (kW)            & (kW)             & (kW) \\ \hline
 Actual Data (2017)            &  --        & --              & 6,353            & -- \\ \hline
 9-year train, 1-year test     & 0.8460     & 18.10          & 6,612            & 85.86\\
 7-year train, 3-year test     & 0.8267     & \textbf{87.54}  & \textbf{7,177}   & 77.80\\
 5-year train, 5-year test     & 0.8134     & 602.3           & 5,482            & 1,055\\
 3-year train, 7-year test     & 0.8308     & 500.7           & 5,713            & 862.1\\
 1-year train, 9-year test     & 0.8332     & \textbf{284.3}  & \textbf{6,061}   & 650.9\\\hline
Actual Data (2019)             &  --        & --              & 6,989            & -- \\ \hline
 1-year train, 11-year test    & 0.8332     & \textbf{346.0}  & \textbf{6,515}  & 690.3 \\ \hline
\end{longtable}

While the mid-summer peak values fall below the actual data for the longer test periods, the peak values in August are relatively aligned with the actual data. To assess the forecast coverage of the actual electric demand values over 5,000 kW, we calculate the RMSE using only those data points. Table \ref{HybridTrainTestB} shows that the RMSE values for over 5,000 kW are slightly higher than the overall RMSE. The model captures the annual growth rates of monthly peak loads and the annual growth rates of overall energy consumption. The model preserves the daily, weekly, and month-to-month trends that occur within each year, or seasonality, of the data. The model realistically represents the electric demand for a full range of weather and occupancy conditions.

\newpage

\newpage

\subsection{University of California Davis}
\label{subsec:AppendixBUCD}

\indent The electric demand data for University of California Davis is shown in Figure \ref{fig:UCDLoad}. A unique attribute of this data is that the peak demand decreases year-after-year despite the increase in full-time equivalent occupancy and the addition of buildings. This school has implemented energy efficiency measures such as equipment replacement, upgrades to the central heating and cooling plant, and lighting retrofits.

\indent In addition to the exogenous variables {\it{Time-Step}, Occupancy, Temperature, Day Category, Class Binary, Cosine Term,} and {\it Sine Term}, the {\it{Energy Use Intensity}} remains after the LASSO selection is performed. An MLR is applied to the exogenous variables after the LASSO selection. The framework is followed with all exogenous variable-based techniques, time series forecast models, and hybrid models resulting in the selection of an MLR model. Table \ref{metrics4B} shows the metrics for 11 different statistical models and highlights the similarity in performance metrics of the exogenous variable-based and hybrid models. 

\begin{longtable}[c]{lcc | ccc}
\caption{Comparative metrics of six-year training and one-year test sets of simple exogenous variable, time-series, and hybrid models. The GAM\textsubscript1 model has a spline cubic regression on the temperature regressor and the GAM\textsubscript2 model has a spline cubic regression on both the temperature and occupancy regressors.} \label{metrics4B}\\
\multicolumn{3}{r}{Six-year training set}   &
\multicolumn{3}{c}{One-year test set}       \\ \hline
Model                           && Adj-$R^2$        & RMSE              & Peak          & Energy Use    \\ 
                                &&                  & (kW)              & (kW)          & (MWh)         \\ \hline
Actual Data                     &&  --              & --                & 41,431        & 211,083       \\ \hline
MLR                             && 0.6033           & 465.3             & 37,630        & 217,021       \\
GAM\textsubscript1              && 0.6099           & 501.5             & 42,520        & 217,339       \\
GAM\textsubscript2              && 0.6323           & 450.5             & 42,851        & 216,892       \\ \hline
ARIMA(4,1,0)                    && 0.9138           & 8,314             & 17,265        & 150,594       \\
ARIMA(2,1,1)                    && 0.9138           & 8,296             & 17,267        & 150,787       \\
SARIMA(0,0,0)(1,1,0)24          && 0.6154           & 5,678             & 23,383        & 180,731       \\
SARIMA(0,1,5)(0,0,1)24          && 0.9235           & 8,032             & 17,180        & 153,512       \\ \hline
MLR+ARIMA(1,0,0)(1,0,0)24       && 0.9351           & 457.9             & 32,251        & 216,957       \\
GAM\textsubscript2+ARIMA(4,1,2)           && 0.6963      & 304.5        & 41,545        & 210,278       \\
GAM\textsubscript1+SARIMA(5,1,2)(1,0,0)24 && 0.6751      & 297.2        & 42,204        & 215,594       \\
GAM\textsubscript2+SARIMA(4,1,0)(1,0,0)24 && 0.6371      & 105.4        & 42,267        & 213,869       \\ \hline
\end{longtable}

\indent While the MLR and GAM\textsubscript2+SARIMA models are both considered using the six-year training set, the MLR model is selected based on the long-horizon forecast. Figure \ref{fig:UCDMLR} shows the MLR six-year forecast. The long-horizon metrics are summarized in Table \ref{metrics5B}. The test RMSE  for the MLR model is 2,169 kW over the six-year forecast and 2,335 kW for the GAM\textsubscript2+SARIMA model. The test maximum is 39,597 kW for the MLR model and 39,730 for the GAM\textsubscript2+SARIMA model, compared to the measured maximum of 42,324 kW. The total energy consumed for the last year of the test is 219,818 MW for the MLR model and 220,571 MW for the GAM\textsubscript2+SARIMA, while the total energy consumed during the last year of measured data is 212,402 MW. The residuals have minimal seasonality present. MLR is the model selected using the framework based on these metrics.

\begin{longtable}[c]{lcc | ccc}
\caption{Comparative metrics of one-year training and six-year test sets the MLR and GAM\textsubscript2+SARIMA models. The GAM\textsubscript2 model has a spline cubic regression on both the temperature and occupancy regressors.} \label{metrics5B}\\
\multicolumn{3}{r}{One-year training set}   &
\multicolumn{3}{c}{Six-year test set}       \\  \hline\hline
Model                           && Adj-$R^2$        & RMSE              & Peak 2023      & Energy Use    \\ 
                                &&                  & (kW)              & (kW)          & 2023 (MWh)    \\ \hline
Actual Data                     &&  --              & --                & 42,324        & 212,402       \\ \hline
MLR                             && 0.6320           & 2,169             & 39,597        & 219,818       \\
GAM\textsubscript2+SARIMA(4,1,0)(1,0,0)24 && 0.7201 & 2,335             & 39,730        & 220,571       \\ \hline
\end{longtable}


\subsection{Clemson}
\label{subsec:AppendixBClemson}

\begin{figure}[htp]
	\centering
        \captionsetup{justification=centering}
	\subfloat[Measured Clemson campus electrical demand data and GAM\textsubscript1 forecast data for a 1.2-year training and a five-year forecast duration.]{\includegraphics[scale=.47]{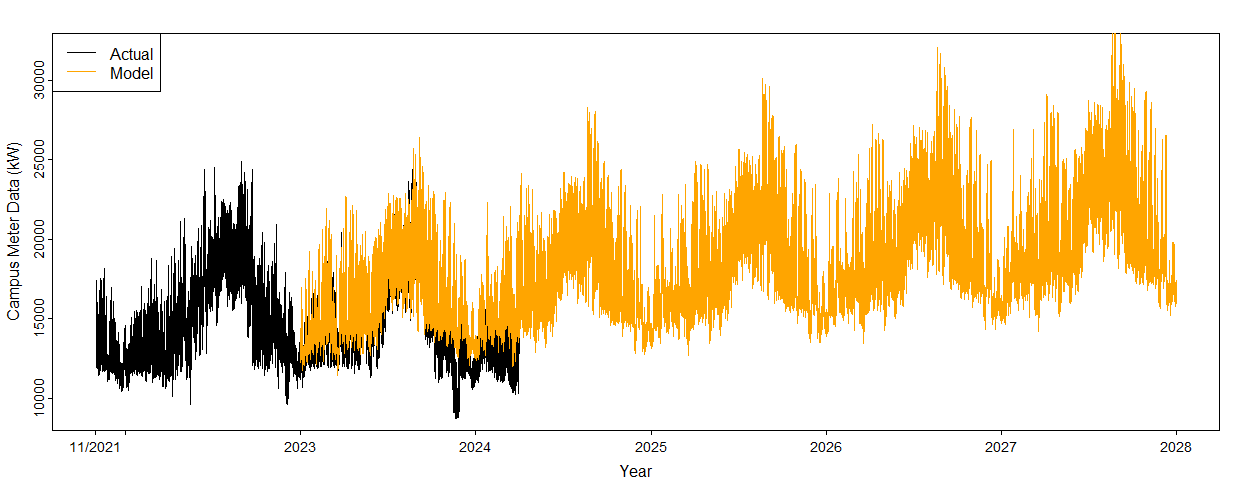} \label{sub:GAM1}} \\
        \subfloat[Measured Clemson campus electrical demand data and GAM\textsubscript2 forecast data for a 1.2-year training and a five-year forecast duration.]{\includegraphics[scale=.47]{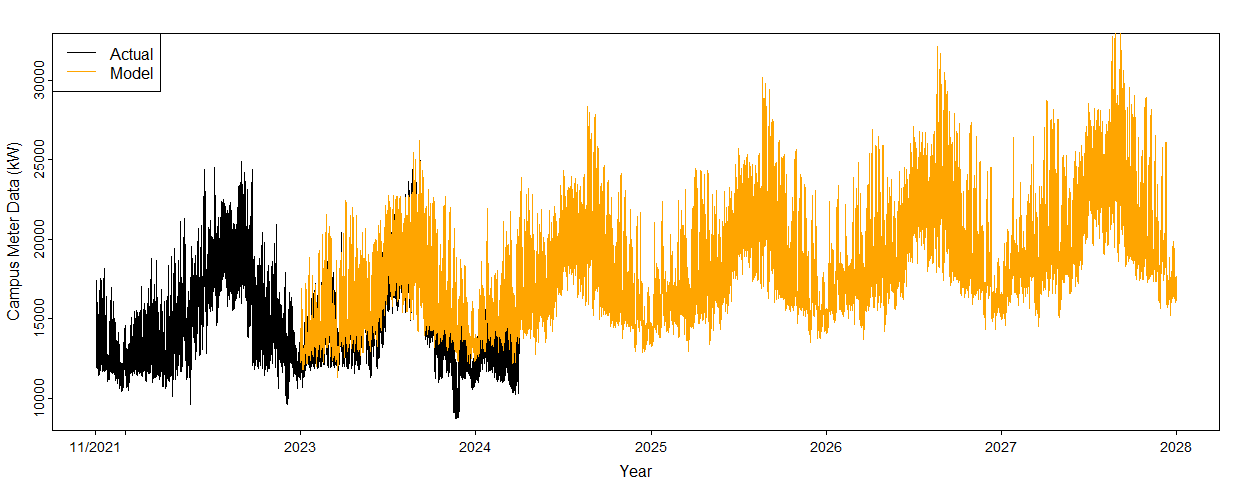} \label{sub:GAM2}}\\   
	\subfloat[Measured Clemson campus electrical demand data and GAM\textsubscript2+ARIMA forecast data for a 1.2-year training and a five-year forecast duration.]{\includegraphics[scale= 0.47]{Clemson_5-year_GAM2+ARIMA.png} \label{sub:GAM2A}}\\
	\caption{Measured Clemson campus electrical demand data and forecast data for selected models for comparison.} 
	\label{fig:ClemsonComp}
\end{figure}

The third data set used to validate the framework is from Clemson, as shown in Figure \ref{fig:ClemsonLoad}. The measured data set has a length of 2.4 years, so the forecast framework was applied for a 1.2-year train and one-year test with the forecast extending over a total of five years, as summarized in Table \ref{Clemsonmetrics1B}. While the measured data for three months of 2024 are shown in Figure \ref{fig:ClemsonLoad} and Figure \ref{fig:ClemsonComp}, the data is disregarded in the test set length due to the energy metrics being compared in full-years only. The exogenous variable outdoor humidity is considered based on the regional significance.

\begin{longtable}[c]{lcc | ccc}
\caption{Comparative metrics of 1.2-year training and one-year test sets of exogenous variable-based, time-series, and hybrid models. The GAM\textsubscript1 model has a spline cubic regression on the temperature regressor and the GAM\textsubscript2 model has a spline cubic regression on both the temperature and occupancy regressors. Bold indicates the best results.} \label{Clemsonmetrics1B}\\
\multicolumn{3}{r}{1.2-year training set}   &
\multicolumn{3}{c}{One-year test set}       \\  \hline\hline
Model                                       & & Adj-$R^2$        & RMSE          & Peak             & Energy Use    \\ 
                                            & &                  & (kW)          & (kW)             & (MWh)         \\ \hline
Actual Data                                 & &  --              & --            & 25,072           & 135,333       \\ \hline
MLR                                         & & 0.726            & 2,591         & \textbf{25,142}  & 193,305       \\
GAM\textsubscript1                          & & 0.815            & \textbf{2,503}& 25,741           & \textbf{192,347}  \\
GAM\textsubscript2                          & & \textbf{0.823}   & 2,665         & 28,404           & 194,116       \\ \hline
ARIMA(5,1,1)                                & & 0.9817           & 3,621         & 12,324           & 103,605      \\
ARIMA(2,1,1)                                & & 0.9778           & 5,048         & 11,164           & 91,100    \\
SARIMA(5,1,1)(1,0,1)24                      & & 0.9862           & 7,468         & 12,368           & 69,900    \\
SARIMA(2,1,1)(1,0,0)24                      & & 0.9809           & 4,938         & 11,408           & 92,068    \\ \hline
MLR+ARIMA(5,1,0)                            & & \textbf{0.9829}  & 2,175         & 16,224           & 116,266   \\
GAM\textsubscript1+ARIMA(5,1,1)             & & 0.8297           & 1,219         & 23,047           & 124,643   \\ 
GAM\textsubscript2+ARIMA(5,1,2)             & & 0.8335           & \textbf{1,045}& \textbf{23,116}  & \textbf{126,168} \\ 
GAM\textsubscript2+SARIMA(5,1,2)(0,0,1)24   & & 0.8412           & 1,505         & 21,911           & 122,135        \\ \hline
\end{longtable}

\indent While the GAM\textsubscript2 and GAM\textsubscript2+ARIMA have higher adjusted $R^2$ values, the peak demands for MLR and GAM\textsubscript1 are closer to the actual data. GAM\textsubscript2+ARIMA, with the annual energy use of the first forecast year being the closest to the measured data, is the selected model. The metrics for the selected GAM\textsubscript2+ARIMA model forecast for the first year are as follows: an RMSE of 1,045 kW, a peak value of 23,116 kW, and total energy of 126,168 MW.  With the test metrics quantifying the quality of only the first year of the forecast, the determination of the best model for subsequent years cannot be established, and the total energy use metric carries weight in the model choice. 

Figure \ref{fig:ClemsonComp} shows the GAM\textsubscript1, GAM\textsubscript2, and GAM\textsubscript2+ARIMA five-year forecasts for visual comparison of the forecast extending four years past the test year.

